\documentclass[a4paper,fleqn,usenatbib]{mnras}

\usepackage[T1]{fontenc}
\usepackage{ae,aecompl}
\usepackage{ulem}

\usepackage{graphicx}	
\usepackage{amsmath}	
\usepackage{amssymb}	
\usepackage{soul}

\title[Scheduling strategies for the ESPRESSO follow-up of TESS targets]{Scheduling strategies for the ESPRESSO follow-up of TESS targets}

\author[L. Cabona et al.]{
L. Cabona,$^{1,2}$\thanks{E-mail: lorenzo.cabona@inaf.it}
P.T.P. Viana,$^{3,4}$
M. Landoni$^{2}$
and J.P. Faria$^{3}$
\\
$^{1}$Università degli Studi dell’Insubria, Via Valleggio 11, I-22100 Como, Italy\\
$^{2}$INAF, Osservatorio Astronomico di Brera, Via E. Bianchi 46 I-23807 Merate (LC), Italy\\
$^{3}$Instituto de Astrof\'{\i}sica e Ci\^{e}ncias do Espa\c{c}o, Universidade do Porto, CAUP, Rua das Estrelas, 4150-762 Porto, Portugal\\
$^{4}$Departamento de F\'{\i}sica e Astronomia, Faculdade de Ci\^{e}ncias, Universidade do Porto, Rua do Campo Alegre, 687, 4169-007 Porto, Portugal
}

\date{Accepted XXX. Received YYY; in original form ZZZ}

\pubyear{2020}

\begin{document}
\label{firstpage}
\pagerange{\pageref{firstpage}--\pageref{lastpage}}
\maketitle

\begin{abstract}
Radial-velocity follow-up of stars harbouring transiting planets detected by TESS is expected to require very large amounts of expensive telescope time in the next few years. Therefore, scheduling strategies should be implemented to maximize the amount of information gathered about the target planetary systems. We consider myopic and non-myopic versions of a novel uniform-in-phase scheduler, as well as a random scheduler, and compare these scheduling strategies with respect to the bias, accuracy and precision achieved in recovering the mass and orbital parameters of transiting and non-transiting planets. This comparison is carried out based on realistic simulations of radial-velocity follow-up with ESPRESSO of a sample of 50 TESS target stars, with simulated planetary systems containing at least one transiting planet with a radius below $4R_{\oplus}$. Radial-velocity datasets were generated under reasonable assumptions about their noise component, including that resulting from stellar activity, and analysed using a fully Bayesian methodology. We find the random scheduler leads to a more biased, less accurate, and less precise, estimation of the mass of the transiting exoplanets. No significant differences are found between the results of the myopic and non-myopic implementations of the uniform-in-phase scheduler. With only about 22 radial velocity measurements per dataset, our novel uniform-in-phase scheduler enables an unbiased (at the level of 1\%) measurement of the masses of the transiting planets, while keeping the average relative accuracy and precision around 16\% and 23\% respectively. The number of non-transiting planets detected is similar for all the scheduling strategies considered, as well as the bias, accuracy and precision with which their masses and orbital parameters are recovered.

\end{abstract}

\begin{keywords}
Planetary systems -- Techniques: radial velocities -- Methods: observational -- Methods: statistical
\end{keywords}



\section{Introduction}
The radial-velocity (RV) follow-up of exoplanet candidates identified using the transit detection method is important to definitively establish their planetary nature, estimate their masses and further refine orbital parameters. It also makes atmospheric studies more informative by constraining the scale height \citep[e.g.][]{BKM2017}. Modelling the internal structure of each exoplanet \citep[e.g.][]{Dorn2015,DBR2018,SCK2018}, and population-level studies, e.g. the characterization of the mass-radius relation \citep[e.g.][]{WRF2016,CK2017,NWG2018,Kanodia2019}, are other applications that benefit from the extra information brought by RV data.

In the next few years, RV follow-up of exoplanet transits will most likely be dominated by observations of TESS [Transiting Exoplanet Survey Satellite, e.g. \citet{R2016}] objects of interest (TOIs). Over the two years of its primary mission, TESS is expected to discover more than 14,000 new transiting exoplanets around almost as many stars \citep{Barclay2018}. The RV measurements required to obtain precise mass measurements even for just a few tens of these planets will easily exceed the many hundreds. Most will be part of concerted efforts by several groups, namely those taking part in the TESS Follow-Up Observing Program (TFOP), with access to large amounts of telescope time. In particular, the ESPRESSO collaboration \citep{Pepe2013,Pepe2014,Pepe2020} plans to devote around 32\% of its Guaranteed Time Observations (GTO) for TOI follow-up, amounting to almost 88 nights distributed across 4 years (N.~C. Santos, private communication).

Often RV measurements for a sample of stars known to host transiting planets are performed in an almost random way, conditional on the target stars being visible at low airmass. More commonly there is some prior planning of the observations, for example to ensure that the RV phase-curves are sampled as uniformly as possible, given the orbital periods inferred from the transit data \citep[e.g.][]{Burt2018}. The most usual stopping criterion for the RV measurements is reaching some relative precision with respect to the transiting exoplanets masses \citep[e.g.][]{Montet2018}. However, in any case, the observations are usually done in a myopic (or greedy) way, i.e. which star is chosen to be observed at a certain time does not take into account all possible scheduling configurations for the future, given the time available and sample of stars to be observed. In principle, this should lead to a less efficient use of available telescope time than non-myopic (also known as batch or block) scheduling. 

Our main objective in this work is then to quantify, using mock but realistic RV simulations, the difference in efficiency, with respect to the information gathered about exoplanet masses and orbital parameters through RV measurements, between different scheduling strategies for the ESPRESSO follow-up of TESS targets. We will consider algorithms whose objective function leads to a sampling of the RV phase-curves of the known transiting planets as uniform as possible, and compare their results with those obtained under an algorithm which just randomly samples the set of target stars that are visible at observation time. The former, henceforth called uniform-in-phase, will be implemented in both a myopic and a non-myopic way.

We start by laying out the procedures used to construct a sample of simulated TOIs, and to generate mock distributions of the ESPRESSO GTO. Next, we describe the scheduling algorithms that will be compared. We then report the results obtained, discuss them, and present our conclusions.

\section{Methods}

\subsection{Stellar sample and simulated planetary systems}

The TESS observing strategy was modelled by \citet{Barclay2018}, in order to identify the approximately 200,000 stars in the TESS Input Catalog Candidate Target List that should be observed at 2-minute cadence. The remaining stars were assumed to be observed at 30-minute cadence in full-frame image data. They then associated zero or more orbiting planets to each star, with specific physical and orbital characteristics, according to measured exoplanet occurrence rates \citep{Fressin2013,DC2015}. Finally, they used the TESS noise model to predict which exoplanets would be detected and their derived properties. It was estimated that TESS would find around 1250 exoplanets in the 2-minute cadence mode, and about 13,100 planets in the full-frame image data. 

A sample of stars for possible ESPRESSO follow-up observations was pre-selected among those stars considered in \citet{Barclay2018} by demanding: a declination in the interval $[-80^{\rm o},\,+30^{\rm o}]$, to ensure extended periods of visibility at low airmass from Paranal; an effective temperature, $T_{eff}$, in the interval $[4000,\,6000$] K, and high surface gravity, $\log g > 4.0$, i.e. only G and K dwarf stars. We then included in our final sample the 50 brightest stars among those pre-selected with at least one orbiting planet with a radius below $4R_{\oplus}$, 3 detected transits and a transit signal-to-noise greater than 10. This final selection step effectively limits our sample to stars with a magnitude, $V$, below 10.5, minimising the RV measurement uncertainty due to photon-noise. It also aligns our sample with a TESS primary science requirement: the estimation of the mass of 50 exoplanets with radius smaller than $4R_{\oplus}$ \citep{R2016}. We ended up with 53 transiting planets orbiting 50 stars, with 3 systems having 2 transiting planets each. We associated to each transiting planet the expected mass, given its radius, obtained using the Forecaster algorithm \citep{CK2017}. The radii were assumed to be known within an uncertainty of $10\%$ (standard deviation), typical of what is expected by combining data from Gaia \citep{Gaia2016,Gaia2018} and TESS \citep{Burt2018}.

In the publicly available from \citet{Barclay2018} catalogue only planets that transit are identified. But, in order to generate realistic simulations of a RV time series, we need to take into account all planets around each star in the sample. Therefore, we added extra orbiting planets to each star, non-detectable by TESS. In order to be coherent with the choice of occurrence rates made in \citet{Barclay2018}, we used for such purpose the occurrence rates published in \citet{Fressin2013}. However, these do not extend to orbital periods long enough to include all planets capable of generating a RV semi-amplitude, $K$, larger than 0.5 m/s, roughly the minimum value we expect our simulated follow-up survey to be sensitive to. This expectation was fully supported a posteriori by the results of the analysis of the simulated RV datasets, we will later describe, as only planets with values for $K$ above $1$ m/s were indeed detected. Although the presence of planets with $K<0.5$ m/s could make the detection of planets with higher values for $K$ more difficult, the effect should be quite small given that we expect that almost always there will be a large difference with respect to $K$ between those extra low-$K$ planets and those that ended up being detected in the systems considered.

Therefore, we first extrapolated the occurrence rates in Table 2 of \citet{Fressin2013} up to orbital periods of 2 years, for radius in the intervals $[2,\,4]$, $[4,\,6]$ and $[6,\,22]$ $R_{\oplus}$, and to 418 days for radius in the interval $[1.25,\,2]$ $R_{\oplus}$. In order to achieve this, we assumed the occurrence rate density, as a function of orbital period, is described by a log-normal distribution \citep[e.g.][]{Fressin2013,WF2015}. The joint posterior probability distribution for the parameters of such function was characterized within each of the four mentioned radius bins, given the occurrence rates provided in Table 2 of \citet{Fressin2013} for the available period bins. The expected values for those log-normal parameters were then used to infer the integrated occurrence rates in the period bins: $[145,\,245]$ and $[245,\,418]$ days in the case of radius between $1.25$ and $2$ $R_{\oplus}$; $[245,\,418]$ and $[418,\,730]$ days in the case of radius between $2$ and $4$ $R_{\oplus}$; $[418,\,730]$ days in the case of radius in the intervals $[4,\,6]$ and $[6,\,22]$ $R_{\oplus}$. These extrapolated occurrence rates can be found in Table \ref{table:t1}, together with the values used from \citet{Fressin2013}. With this extrapolation, we are able to take into account all planets, with an orbital period smaller than 2 years, that are capable of inducing a RV signal with $K>0.5$ m/s, given their expected mass as estimated using the Forecaster algorithm \citep{CK2017}.

\begin{table*}
\caption{Average number of planets per star per radius and period bin (in percent) from \citet{Fressin2013}. Inside square brackets are extrapolated values by assuming that, inside each radius interval, the occurrence rate density is described by a log-normal function of the orbital period.}
\centering
\begin{tabular}{c c c c c c}
\hline\hline
\begin{tabular}{@{}c@{}}Period Range \\(days) \end{tabular} &  \begin{tabular}{@{}c@{}} Giant \\ (6 - 22 $R_{\oplus}$) \end{tabular} & \begin{tabular}{@{}c@{}}  Large Neptunes\\ (4 - 6 $R_{\oplus}$) \end{tabular}   & \begin{tabular}{@{}c@{}} Small Neptunes\\(2 - 4 $R_{\oplus}$) \end{tabular}  & \begin{tabular}{@{}c@{}}Super-Earths\\(1.25 - 2 $R_{\oplus}$)\end{tabular}  & \begin{tabular}{@{}c@{}} Earths \\ (0.8 - 1.25 $R_{\oplus}$) \end{tabular}  \\
0.8-2.0 & 0.015 & 0.004 & 0.035 & 0.17 & 0.18\\
2.0-3.4 & 0.067 &0.006 &0.18 &0.74 &0.61\\
3.4-5.9 & 0.17 & 0.11 & 0.73 & 1.49 & 1.72\\
5.9-10 & 0.18 & 0.091 & 1.93 & 2.90 & 2.70\\
10.0-17.0 & 0.27 &  0.29 & 3.67 & 4.30 &  2.70\\
17.0-29.0 & 0.23 & 0.32 & 5.29 & 4.49 & 2.93\\
29.0-50.0 & 0.35 & 0.49 & 6.45 & 5.29 & 4.08\\
50.0-85.0 & 0.71 & 0.66 & 5.25 & 3.66 & 3.46\\
85.0-145.0 & 1.25 & 0.43 & 4.31 & 6.54 & -\\  
145.0-245.0 & 0.94 & 0.53 & 3.09 & [0.91] & -\\
245.0-418.0 & 1.05 & 0.24 & [1.89] & [0.35] & -\\
418.0-730.0 & [0.91] & [0.12] & [0.75] & - & -\\
\hline\hline
\end{tabular}
\label{table:t1}
\end{table*}

The number of planets we associate with each star, within the radius-period bins identified in Table 2 of \citet{Fressin2013} plus those with extrapolated occurrence rates, was then randomly drawn from a Poisson distribution with mean 0.92 (expected number of planets across all such bins). If the number obtained was greater than the number of transiting planets in the system, the radius-period bins where the extra planets are located were randomly drawn from the full radius-period bin distribution taking into account the respective occurrence rates. Then, a specific radius and period was randomly drawn for each extra planet inside the associated bin, assuming a log-normal distribution (the same type that was considered in the radius-period bin occurrence rates extrapolation). For each extra planet, the radius and orbital period drawing procedure is repeated until the transit signal-to-noise is lower than 10, or the period found is greater than twice the timespan of the scheduled TESS observations of the sector where the star is located. This ensures that any extra planet associated with the stars in our sample would not have been detected in the simulations of TESS observations made by \citet{Barclay2018}. 

Since planets with radius above 4 $R_{\oplus}$ can have $K>0.5$ m/s even with orbital periods greater than 2 years, we randomly added extra planets with orbital period between 2 and 10 years. For this we used the occurrence rates in \citet{HZW2019}, respectively $0.24$ and $0.15$ in the radius ranges $[4.5,\,9.5]$ and $[9.5,\,15.7]$ $R_{\oplus}$. We again associated to each extra planet the expected mass, given its radius, obtained using the Forecaster algorithm \citep{CK2017}.

We ended up with 50 extra planets, distributed across 35 systems (only one of which has 3 extra planets). Their orbital eccentricities, $e$, were then randomly drawn from a Beta distribution with parameters $\alpha = 1.03$ and $\beta = 13.6 $ following \citet{Kipping2014}. We kept the assumption of \citet{Barclay2018} that all planets in any system are co-planar, and set the inclination of all systems to $90^{\rm o}$ in order to enable a more direct comparison between true and estimated planetary masses. At this point, we determined whether each possible pair of planets in any given system is Hill stable, by finding if the following inequality is true \citep{Gladman1993}:

\begin{equation}
\left(\mu_{1} + \mu_{2} \frac{a_1}{a_2} \right) \left(\mu_{1} \gamma_{1} + \mu_{2}\gamma_{2} \sqrt{\frac{a_{2}}{a_{1} }} \right)^{2} >\alpha^{3} + 3^{4/3}\mu_{1}\mu_{2}\alpha^{5/3},
 \end{equation}
where $\mu_{i}$, $a_{i}$ and $e_{i}$ are respectively the ratio between the planet mass and the mass of the star which it orbits, the orbital semi-major axis, and the orbital eccentricity, with $\alpha=\mu_{1}+\mu_{2}$ and $\gamma_{i}=\sqrt{1-e_{i}^{2}}$, for each planet $i=\{1,\,2\}$ in the pair being considered. All the systems found to contain Hill unstable pairs of planets were re-simulated, keeping the number of extra planets but randomly re-drawing their radius and orbital parameters, until every simulated planetary system only contained Hill stable pairs. In the process, we actually found that the pair of transiting planets in system with identification number 304142124 is not Hill stable. In order to minimally change the catalogue published by \citet{Barclay2018}, while ensuring this planet pair becomes Hill stable, we just decreased the eccentricity of the outer transiting planet from $0.15453$ to $0.142$.

Finally, for both transiting and non-transiting planets, the mean anomaly $M_0$ at the time $t_0$ (when we start our scheduler), and the argument of periastron, $\omega$, were randomly drawn from a uniform distribution between 0 and $2\pi$. With this it becomes possible to compute the overall planetary contribution to the RV time series for each star.

\begin{figure}
\includegraphics[width=\columnwidth]{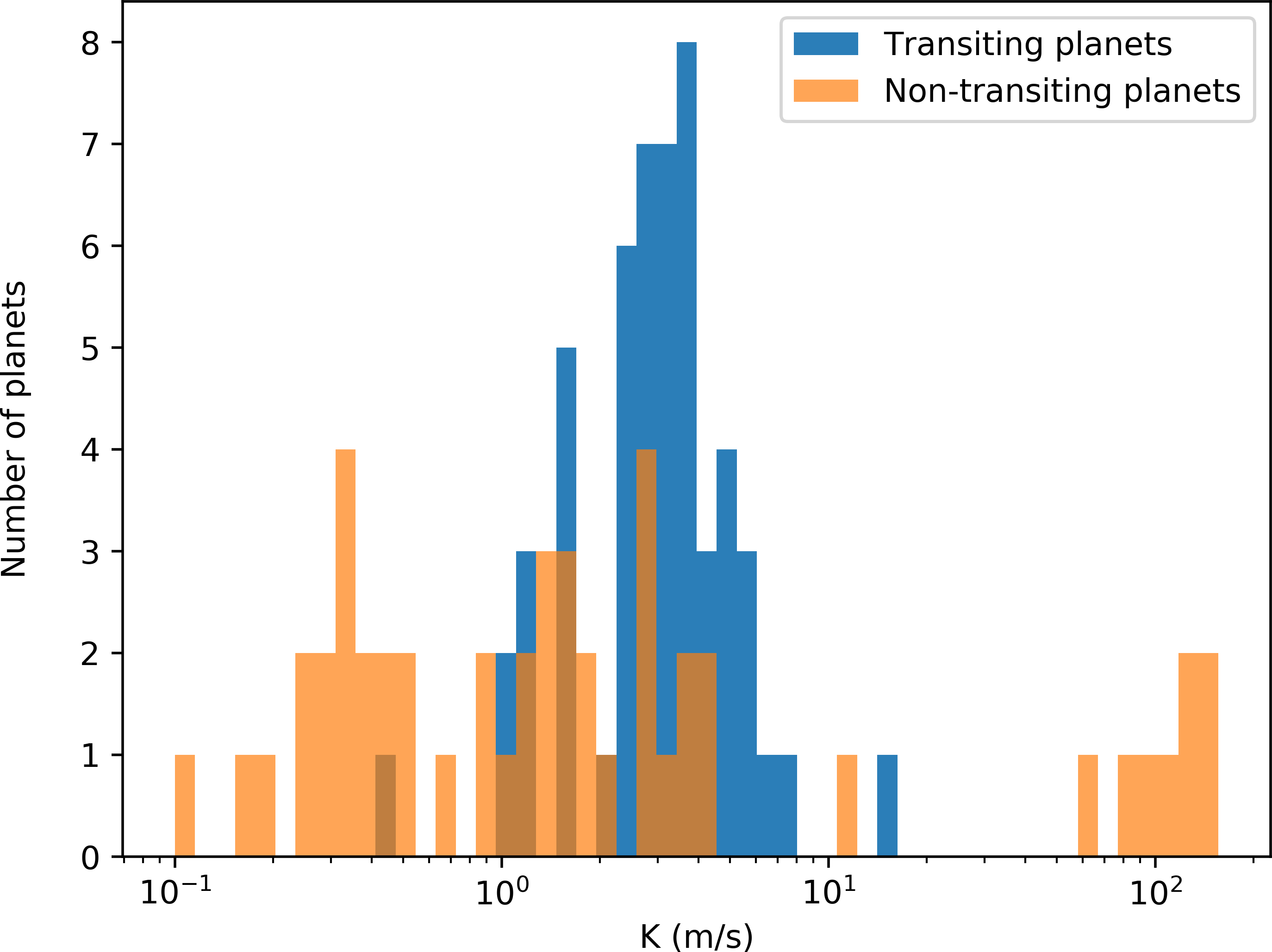}\vspace{0.1cm}
\includegraphics[width=\columnwidth]{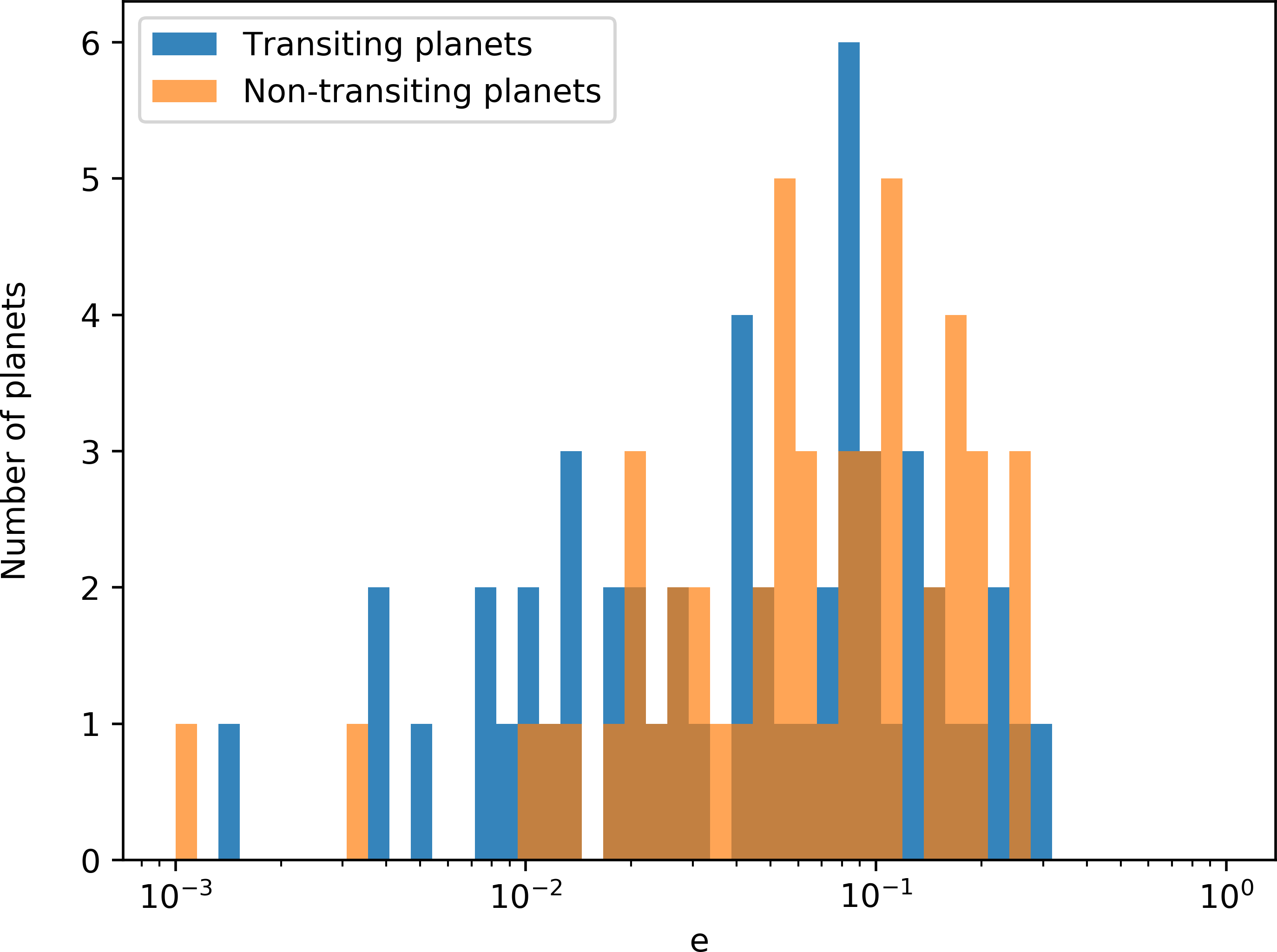}\vspace{0.1cm}
\includegraphics[width=\columnwidth]{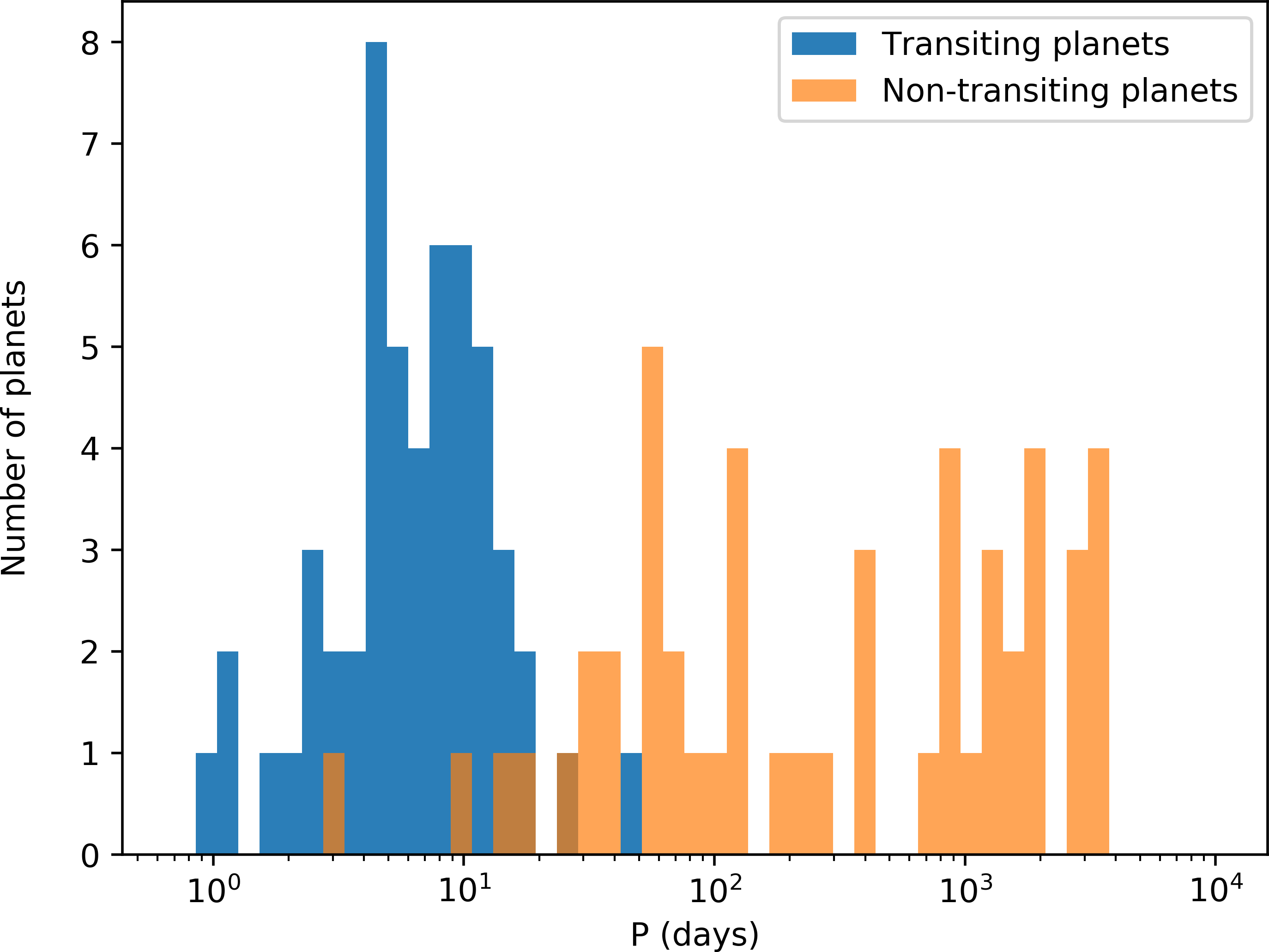}
\caption{From upper to lower panel, distributions of RV semi-amplitudes, $K$, orbital eccentricities, $e$, and periods, $P$, for the transiting (blue) and non-transiting (orange) planets.}
\label{fig:i1}
\end{figure}

Figure \ref{fig:i1} shows the distributions for $P$, $K$ and $e$, for the simulated transiting and non-transiting planets. A detailed description of the properties of every planet in our simulation is provided in a machine readable table, with a summary shown in Table \ref{table:t2}.

\begin{table}
\caption{Summary of the properties of all stars and planets in the sample.}
\centering
\begin{tabular}{l c}
\hline\hline
Column&Property\\
\hline
1  & Sample ID number of star \\
2  & TESS Input Catalog ID number of star \\
3  & Right ascension 2000 (in degrees)\\
4  & Declination 2000 (in degrees) \\
5  & V-band magnitude\\
6  & Stellar effective temperature (in K)\\
7  & Stellar radius (in $R_{\sun}$)\\
8  &  Stellar mass (in $M_{\sun}$)\\
9  & Number of TESS sectors the star is observed in\\
10  & One-hour integrated noise level of the star (in ppm)\\
11  & Stellar flicker, $F_8$ (in ppt)\\
12  & Stellar jitter, $\sigma_{\rm act}$ (in m/s)\\
13  & Stellar rotation period, $P_{\rm rot}$ (in days)\\
14  & Systemic velocity (in m/s)\\
15  & Planet orbital period (in days) \\
16  & Planet orbital eccentricity\\
17  & Argument of Periastron (in rad)\\
18  & Time of periastron passage (in BJD)\\
19  & Time of transit (in BJD)\\
20  & Mean anomaly at time $t_0$ (in rad)\\
21  & Planet radius (in $R_{\oplus}$)\\
22  & Planet mass (in $M_{\oplus}$)\\
23  & Radial velocity semi-amplitude (in m/s)\\
\hline
\end{tabular}
\label{table:t2}
\end{table}

\subsection{ESPRESSO GTO simulations}

The ESPRESSO GTO consists of 273 nights during 4 years, and began on the 1st of October 2018\footnotemark\footnotetext{\url{https://www.eso.org/sci/observing/policies/gto_policy.html}}. Exoplanetary science occupies 80\% of the time, 10\% is allocated to fundamental constants time-variability studies and 10\% is discretionary time at the disposal of the ESPRESSO consortium \citep{Pepe2013,Pepe2020}. The total amount of time available for exoplanetary science is in turn divided as follows: 30\% for exoplanetary atmospheric characterization; 30\% for TOI follow-up; 40\% for a RV survey. We simulated the scheduling of ESPRESSO GTO observations from the 1st of October 2019 until the 30th of September 2022, i.e. only for 3 years. Furthermore, we assumed that on the 1st of October 2019 all our TOIs would have been observed and characterized by TESS.

The 80\% of the ESPRESSO GTO dedicated to exoplanetary science consists of close to 55 half-nights each semester. We randomly spread them in such a way as to mimic the ESPRESSO GTO distribution in ESO periods 102 and 103, the only known at the time of writing, including aggregation of some half-nights into full nights. Each full day is divided into 60 observation slots, all with a duration of 24 minutes (15 as integration time plus 9 for overheads), but due to seasonal variation, each astronomical night will have a different number of observation slots associated. The integration time was defined to be 15 minutes in order to average out the RV variability induced by stellar oscillations in the G and K dwarf stars we are considering \citep[e.g.][]{Dumusque2011}. We will only consider observations slots with an associated airmass not greater than 2.0. Thus, taking into account the magnitude and temperature ranges for the 50 stars we are considering, respectively, $[6.69,10.37]$ and $[4408,5978]$ K, the ESPRESSO ETC (Exposure Time Calculator) \footnotemark\footnotetext{\url{https://www.eso.org/observing/etc/bin/gen/form?INS.NAME=ESPRESSO+INS.MODE=spectro}} estimated RV variability due to photon-noise will range from $0.1$ to $0.5$ m/s, under normal atmospheric conditions. The average value is close to $0.3$ m/s across all observational slots for which RV simulations were performed.

We further assumed that exoplanetary atmospheric characterization takes precedence, given that they are performed during transit and thus are time-critical. For each semester we thus first randomly sampled, with repetition, the ESPRESSO consortium target list for this type of study, until 30\% of the available time was reached. Each scheduled transit observation is composed of enough sequential observational slots to cover the time interval from one hour before the transit starts until one hour after the transit ends. Some of the half and full nights allocated to exoplanetary atmospheric characterization are not completely filled with these type of observations and thus the remaining slots are available for TOI follow-up and the RV survey. We repeated this procedure 10 times, obtaining 10 different distributions for the 80\% of the ESPRESSO GTO dedicated to exoplanetary science. These simulations yielded between 2563 and 2628 (24-minute) slots that can be used for TOI follow-up and the RV survey. Among these we decided to schedule a fixed number of 1102 slots for TOI follow-up, which we assume take precedence over the RV survey. This number is very close to the fraction that can be used for TOI follow-up, i.e. 30\% of the total number of slots associated with each GTO realization. Although we could have let that number vary with each GTO simulation, we decided to fix it to the mean averaged over all simulations so that the results could be more easily compared.

\subsection{RV simulations}

Stellar activity also induces variations in the radial velocity of a star \citep[e.g.][]{Korhonen2015,Dumusque2016,Cameron2018,Cegla2019}. Their overall amplitude, $\sigma_{\rm act}$, was determined, for all stars in our sample, by randomly drawing from a Normal distribution with mean given by Equation 4 in \citet{Cegla2014}, and a standard deviation of $0.4$. The mean reproduces the observed correlation between RV variability and a measure of stellar flicker, $F_8$, for stars with low levels of activity, while the value of $0.4$ is suggested by Figure 6 in \citet{Cegla2014}. The flicker parameter, $F_8$, is determined using Equation 2 in \citet{TSC2018}, which depends on stellar mass, effective temperature and $\log g $, information we have for all the TESS target stars we consider. The assumed value of $\sigma_{\rm act}$ for each star in our sample can be seen in Figure \ref{fig:d_wn}, in the Appendix.

The RV variations induced by stellar activity have been shown to be well modelled as a Gaussian Process, i.e. their joint probability distribution is assumed to be a multivariate Normal with a number of dimensions equal to the number of RV measurements under analysis, and in particular as a Gaussian Process with a mean of zero and a covariance matrix, $\Sigma$, with entries calculated using a quasi-periodic covariance function or kernel \citep[e.g.][]{Haywood2014,Faria2016,Rajpaul2017,Angus2018}. Nevertheless, as we will later see, none of the three scheduling strategies under study relies on the assumed model for the stellar activity induced RV variations to decide on the best schedule, thus any changes to such model should have little impact on the relative outcomes of those strategies. In order to test this, we performed RV simulations where the impact of stellar activity was assumed to be the result of either a quasi-periodic Gaussian Process or Gaussian white noise, i.e. randomly and independently generated (as a function of time) from a Normal distribution with a constant mean (zero) and standard deviation ($\sigma_{\rm wn}$, equal to $\sigma_{\rm act}$ in our case). Note that Gaussian white noise is equivalent to a Gaussian Process with a covariance matrix, $\Sigma$, whose entries are zero everywhere except in its diagonal, where they are equal to the square of the assumed standard deviation (i.e. the variance).

The assumption of a quasi-periodic kernel gives rise to entries in the covariance matrix, $\Sigma$, of the form
\begin{equation}\label{eq:qp-kernel}
\Sigma_{ij} = \eta_1^2 
    \exp\left[ - \frac{(t_i - t_j)^2}{2\eta_2^2} - \frac{2\sin^2\left(\frac{\pi (t_i-t_j)}{\eta_3}\right)}{\eta_4^2} \right] + s^2 \delta_{ij} \:\text{,}
\end{equation}
  \noindent
where $\eta_1$, $\eta_2$, $\eta_3$, and $\eta_4$ are parameters that can be interpreted as the amplitude, timescale of decay, periodic timescale, and level of high-frequency variability (within the periodic timescale) of the RV variations. Because these are caused by stellar active regions, whose appearance and disappearance along the line-of-sight is modulated by the stellar rotation period, $\eta_3$ should be close to its value. The parameter $s$ is usually known as jitter, and accounts for (apparent) non-correlated variability. This can arise, for example, through sampling with a much lower frequency than that associated with correlated RV variability induced by one or more of the many physical process involved in stellar activity.

For each star, the value of $s$ was randomly drawn with equal probability from the interval $[0.1\,{\rm m/s},\,\sigma_{\rm act}/2]$, while the value for $\eta_1$ was assumed to be the square-root of $\sigma_{\rm act}^2$ minus $s^2$. The assumed values of $s$ and $\eta_1$ for each star in our sample can be seen in Figure \ref{fig:d_gp}. We also set $\eta_3$ equal to the rotation period of each star. This was fixed using the following procedure. First, we selected all stars with an effective temperature within $[4000,\,6000$] K and $\log g > 4.0$, included in the catalogue of 34,000 Kepler main sequence stars assembled by \citet{McQuillan2014}. Then, for each star, we identified its nearest 100 neighbours in the plane defined by effective temperature and $\log g$, determined the mean and standard deviation of the distribution of the rotation periods for those 100 stars, and randomly drew a value from a Gaussian with the derived mean and standard deviation. Finally, we associated to the star $i$ under consideration the measured rotation period, $P_{{\rm rot}\,,i}$ (and its associated uncertainty, $\sigma_{{\rm rot}\,,i}$) in the sample of 100 stars that is closest to the value previously drawn from the Gaussian. The rotation period assumed for each star in our sample can be found in Table \ref{table:t2}. For all stars in our sample, the values of $\eta_2$ were randomly drawn from a log-uniform distribution truncated at $\eta_3$ and $5\eta_3$, while the values of $\eta_4$ were randomly drawn from a Gaussian distribution with mean of $0.7$ and standard deviation of $0.05$. This ensures the values thus obtained are typical of those inferred from the analysis of RV data \citep[e.g.][]{Faria2016,Morales2016,Cloutier2017,Faria2020}.

We consider two other contributions to the variability of the RV time-series, which we assume can be characterized as Gaussian white noise: one due to photon-noise, $\sigma_{\rm ph}$, which was calculated using the ESPRESSO ETC specifically for each star according to its magnitude, effective temperature and airmass at the time of observation; and another due to the RV variability induced by instrumental-noise, $\sigma_{\rm ins}$, which we assumed to be constant and equal to $0.1$ m/s \citep{Pepe2014,Pepe2020}. Therefore, the full covariance matrix, $\Sigma_v$, associated with the multivariate Normal that describes the stochastic behaviour of each RV time-series becomes equal to the covariance matrix, $\Sigma$, associated with the specific Gaussian Process used to describe stellar activity induced RV variability (either quasi-periodic or white noise), to whose diagonal the squares of both $\sigma_{\rm ph}$ and $\sigma_{\rm ins}$ are added.

We also associated to every star a systemic RV relative to the centre of mass of the system, $v_{\rm sys}$, drawn from a random uniform distribution between $-100$ to $100$ m/s, roughly the observed range for stars in the solar neighbourhood \citep[e.g.][]{KSB2017}. Thus, the RV time series associated with each star were generated based on the following model:
  
\begin{equation}
v_{\rm r}(t) = v_{\rm sys} + \sum\limits_{\rm i=1}^{\rm n_p} v_{\rm r,i}(t) + \epsilon(t)
\label{keplerian1}
\end{equation}
with
\begin{equation}
 v_{\rm r,i}(t)=K_{\rm i}\{\cos [\phi_{\rm i}(t)+\omega_{\rm i}]+e_{\rm i}\cos(\omega_{\rm i})\}
 \label{keplerian2}
\end{equation}
\noindent
\begin{equation}
\epsilon(t)\sim N\left(0,\Sigma_v\right)
\label{keplerian3}
\end{equation}
where ${\rm n_p}$ is the number of planets orbiting the star, $K_{\rm i}$ is the RV semi-amplitude, $\omega_{\rm i}$ is the argument of periastron, $e_{\rm i}$ is the orbital eccentricity, and $\phi_{\rm i}(t)$ is the true anomaly as a function of time, $t$, calculated from the other orbital parameters \citep[e.g.][]{Perryman2018}, all with respect to planet i. Thus, we neglect any gravitational interactions between orbiting planets when calculating the instantaneous RV for every star. 
\subsection{Scheduling strategies}

\begin{figure*}
\includegraphics[width=\linewidth]{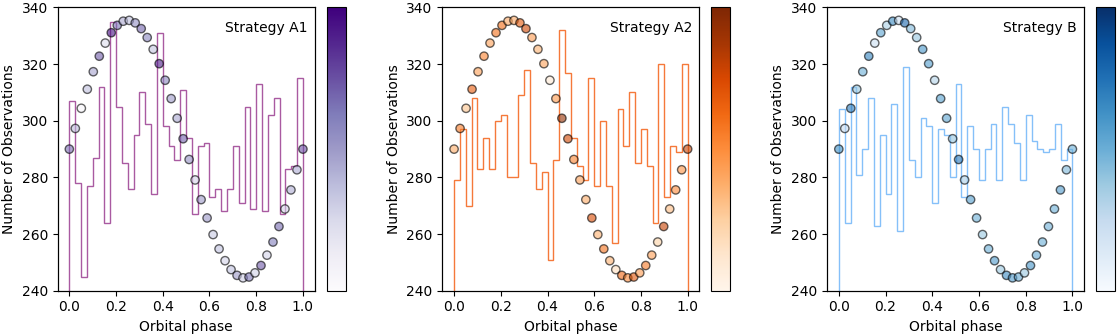}
\caption{Total number of RV observations scheduled, averaged over 10 simulations per strategy, as a function of the orbital phase of the transiting planets when each system is observed, for the three scheduling strategies, A1 (left panel), A2 (central panel ) and B (right panel). Both the vertical axis and the colour gradient indicate the number of RV observations per bin. Phase zero for each planet corresponds to its crossing of the line-of-sight.}
\label{fig:i3}
\end{figure*}

\begin{figure*}
\includegraphics[width=\linewidth]{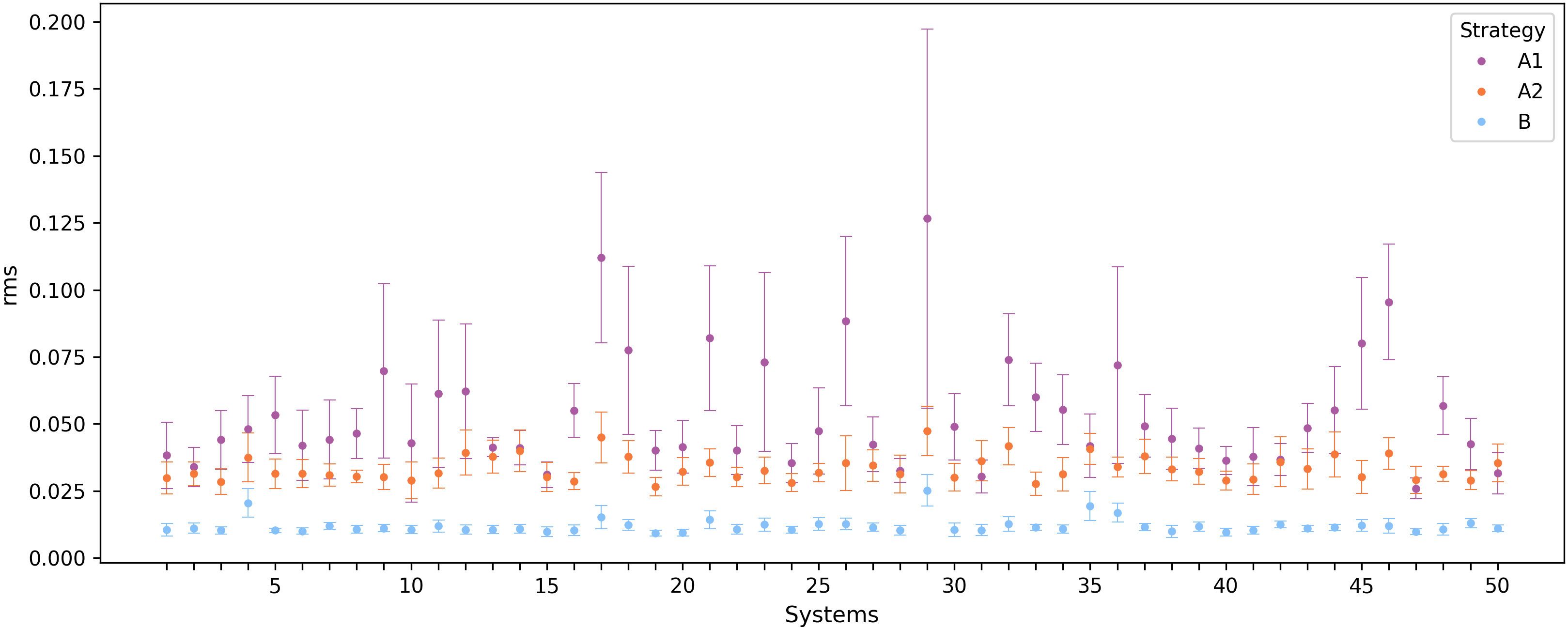}
\caption{Root mean square (rms) of the difference between the simulated orbital phase coverage of transiting planets and perfectly uniform phase sampling, averaged over all such planets in each system and the simulated datasets, per strategy: A1, magenta; A2, orange; B, cyan. The bars represent the standard deviations of the rms with respect to each set of 10 simulations. The systems are identified by an incremental number where 1 corresponds to the lowest TESS ID number and 50 to the highest TESS ID number in our sample.}
\label{fig:i4}
\end{figure*}

We will consider three different scheduling strategies. Two of them, labelled A, are myopic, i.e. the best schedule is defined sequentially in time. In strategy A1, the star chosen to be observed at any given time is randomly drawn from all stars in the sample which can be observed at that time, at an airmass equal or smaller than 2, and with a Moon separation greater than 30 degree, henceforth known as the observability constraint. In strategy A2, this sub-sample of stars is further restricted to the stars that have a smaller number of observations than those associated with the sample star with the largest number of allocated observations at previous times, henceforth known as the equalizing condition. Imposing the second condition leads to a more even distribution of the observational slots between the sample stars. 

However, in the case of strategy A2, we also want the sampling of the RV phase-curves of the known transiting planets to be as uniform as possible, i.e. to ensure as close as possible uniform-in-phase sampling. This is achieved through the maximization of the following objective function, capable of measuring the overall dispersion of points in a given interval,

\begin{equation}
f(\{x_{\rm i}\})\equiv\left\{\sum\limits_{\rm i=1}^{\rm 1102} [d(x_{\rm i})]^{-q}\right\}^{-1/q}
\end{equation}
where $d(x_{\rm i})$ is the time distance between the observation $x_{\rm i}$ and its nearest neighbour in the orbital phase-space of the transiting planet targeted by the observation (including across the phase-space boundary), as a fraction of the orbital period of such planet. When more than one transiting planet exists around a star, $d(x_{\rm i})$ equals the sum of the distances with respect to all transiting planets in the system, which favours the observation of stars for which multiple transiting planets are known. We assume the orbital period and mid-transit time for each transiting planet to be perfectly known a priori. In a real application, this would mean fixing them to e.g. their expected values given the TESS data. 

In the context of strategy A2, the best schedule is then also constructed sequentially in time. First, the stars that fulfil both the observability constraint and the equalizing condition are identified. The star chosen to be observed among these will be the star with the smallest number of allocated observations at previous times. If several stars share this number of previous observations, then the star chosen to be observed is that which leads to the maximization of the objective function provided, $f(\{x_{\rm i}\})$. This same criterium is applied to all stars that satisfy the observability constraint in the situation where no stars fulfil the equality condition (i.e. all have the same number of allocated observations at previous times). However, these rules are applied only at those times for which all stars that satisfy the observability constraint and the equalizing condition have already been observed at least once, else the star chosen to be observed is randomly drawn among those that have not yet been so. As a result of the this procedure, all 10 simulated schedules, according to strategy A2, associate between 17 and 24 observational slots to each star. In contrast, strategy A1 always leads to some stars being observed only a few times, the minimum ranging from 2 to 8 across the 10 simulations, while some other stars end up being slotted for observation as many as 39 to 49 times.

The third strategy, labelled B, is non-myopic. In this case, the aim is to compare all possible schedules, across the full time-span of 3 years, and then choose that which maximizes the objective function, $f(\{x_{\rm i}\})$. Given the form taken by such function, this procedure leads to what is known as $L^q$ relaxation of the points in the design space, in our case the orbital phase-space of each planet. It yields a nearly optimal approximation to the maximin solution to the problem \citep[e.g.][]{Pronzato2017}. The larger the $q$, the better should be this approximation. But then the objective function becomes increasingly localised (in the space of all possible scheduling configurations) and it is more difficult to find the region where the function is maximised. After extensive testing we decided to use $q=2$ (also in the case of strategy A2 to allow for easier comparison between myopic and non-myopic scheduling).

The maximin solution is a classical example of a space-filling strategy \citep[see e.g.][for a review]{Pronzato2017}. In our case, it corresponds to finding the schedule that maximizes the sum over all stars of the minimum (time) distance, normalized as a fraction of the orbital period(s) of the known transiting planet(s) around each star, between any observation and all others of the same star. An alternative classical space-filling strategy is the minimax solution. In this case, the objective would be to find the schedule that minimizes the sum over all stars of the maximum distance (as defined before) between any observation and all others of the same star. However, the maximin solution is computationally faster to find, because it only requires the calculation of distances between neighbouring observations in the orbital phase-space of each planet. Whereas finding the minimax solution would require the calculation of the distances between all observations with respect to each planet \citep[e.g.][]{Pronzato2017}. Nevertheless, we also implemented an algorithm to identify the minimax solution, and found it leads to schedules very similar to those obtained using strategy B.

Given the large number of time slots available for scheduling and the fact that the stars considered are observable during most of any given year, the number of possible scheduling configurations is very large. Therefore, it is impossible to compare the values the objective function takes for all such configurations. As a result, we used the acebayes R package\footnotemark\footnotetext{\url{https://cran.r-project.org/web/packages/acebayes}} \citep{OWA2017} to find the schedule that maximizes the objective function. This is done via an approximate coordinate exchange (ACE) algorithm, where a sequence of conditional one-dimensional optimisation steps are used, as described in \citet{OW2017}. In our case, the objective function depends on the stellar label and the slot time, which will hence be our coordinates. Each schedule, or design, can be viewed as a collection of points in this two-dimensional space. The search for the maximum of the objective function then proceeds through the sequential change of the coordinates of each point in a given initial schedule. In the case of the stellar label coordinate, a change occurs when it is found that, for the time slot associated with a particular design point, there is another star for which the objective function attains a higher value and each star in the sample continues to be observed within the pre-specified minimum number of times. In the case of the time coordinate, a change occurs if there is another time slot for which the objective function reaches a higher value, among those which are not yet associated with a star and for which the observability constraint is obeyed. The search for such optimal time slot  is performed by first approximating the objective function with respect to observation time, for the star associated with the design point under consideration, through a Gaussian process (within acebayes), and then by identifying the time for which the objective function is maximized.

The initial schedule for strategy B is created randomly, with the only conditions being that the observability constraint is obeyed and each star in the sample is observed at least the pre-specified minimum number of times. The closer the latter is to the average number of available observational slots per star, in our case $1102/50\simeq22$, the harder it is for the ACE algorithm to optimize the schedule, and the smaller will be the value of the objective function at the end of the optimization process. This means that there is a trade-off between ensuring an (almost) equal number of observations per star and an optimized sampling of the orbital phase-space of each transiting planet. Somewhat arbitrarily, we set the required minimum number of observational slots per star to be 20. If it was much smaller, there would be significant variations in the accuracy and precision with which the mass and orbital parameters of each transiting planet would be recovered. 

In our case, each run of the ACE algorithm goes through a sequence of $2\times1102=2204$ conditional optimisation steps. In order to consolidate the best schedule, we re-run 100 times the ACE algorithm within acebayes, using the output of each ACE run as input to the following one. As the runs progress, we keep track of the objective function value, and choose the final best design (which is not necessarily the last) as that with the highest associated value for the objective function.

In Figure \ref{fig:i3} we show how many RV observations are scheduled, for all 10 simulations per strategy, as a function of where each observed planet is in the respective phase-curve. This is equivalent to seeing the phase-curves in overlap, and as expected all scheduling strategies lead to almost uniform distributions. However, this hides significant differences in the phase-space distribution of RV observations between transiting planets. In particular, as expected, strategy A1 leads to the most irregular phase-curve coverage per planet and dataset, followed by strategy A2. This can be clearly seen in Figure \ref{fig:i4}. This shows the root mean square (rms) of the difference between the simulated orbital phase coverage of transiting planets and perfectly uniform phase sampling, averaged over all such planets in each system and the associated 10 simulated datasets per strategy. This difference is obtained, for each planet and simulation, by summing the squares of the distances between the sorted orbital phases (between 0 and 1), with each distance subtracted by the inverse of the number of RV measurements (which is the distance between orbital phases under perfectly uniform phase sampling). As expected, the orbital phase coverage is much consistently (low standard deviation) closer to uniform (low average) in the case of strategy B than in the case of the other two strategies, especially A1.

In practice, our assumption in the case of strategy B that the ESPRESSO GTO schedule can be known a priori for the full 3 years is unrealistic. ESO will only inform the ESPRESSO consortium of its schedule for each semester close to its beginning. Therefore, a more realistic implementation of strategy B would require re-scheduling every 6 months the remaining time for the completion of the 3 years. This should not have a significant impact in the expected efficiency with which information is recovered about planets properties through the implementation of strategy B. This is because what is expected to happen within each semester, as ESO relays the information about the available observational slots, is just an effectively random re-shuffling of their position within the semester. Thus, the expected information gain guiding strategy B should remain essentially the same. A more realistic implementation of this strategy should only suffer from some loss of coherence around the start/end of each semester, the more so the smaller the orbital periods of the systems scheduled to be observed at those times. This near-randomization of the scheduler at such a small fraction of the available time should have a very small impact on the expected information gathered through strategy B. Given the considerable amount of extra computing time required to simulate a re-scheduling every 6 months, we decided to implement strategy B in the more simplified manner previously presented.

\section{Results and discussion}

\subsection{Bayesian analysis}

\begin{table}
\caption{Prior distributions for the parameters in the RV meta-model.}
\centering    
\begin{tabular}{lc}
\hline\hline
Transiting planets & \\
\hline\hline
$P$ [days] & $\mathcal{G}\,(P_{{\rm TESS},i},\,0.001)$ \\
\hline
One per system & \\
\hline
$e$ & $\mathcal{HG}\,(0, 0.32)$ \\
\hline
Two per system & \\
\hline
$e$ & $\mathcal{HG}\,(0, 0.083)$ \\
\hline\hline
Non-transiting planets & \\
\hline\hline
${\rm n_{nt}}$ & $\mathcal{U}\,(0, 5)$ \\
$P$ [days] & $\mathcal{J}\,(1 \,, 10000)$ \\
$e$ & $\mathcal{B}\,(0.867, 3.03)$ \\
$M_0$ [days] & $\mathcal{U}\,(0, 2\pi)$ \\
\hline\hline
All planets & \\
\hline\hline
$K$ [m/s] & $\mathcal{MJ}\,(1, 1000)$ \\
$\omega$ [rad] & $\mathcal{U}\,(0, 2\pi)$ \\
\hline          
$ v_{\rm sys}$ [m/s] & $\mathcal{U}\,({\rm RV}_{{\rm min},i},\,{\rm RV}_{{\rm max},i})$ \\
$\eta_1$ [m/s] & $\mathcal{J}\,(0.1, {\rm RV}_{{\rm max},i}-{\rm RV}_{{\rm min},i})$ \\
$\eta_2$ [days] & $\mathcal{J}\,(P_{{\rm rot},i}, 5P_{{\rm rot},i})$ \\
$\eta_3$ [days] & $\mathcal{G}\,(P_{{\rm rot},i},\,5\sigma_{{\rm rot},i})$ \\
$\eta_4$ & $\mathcal{J}\,(1/e,\,e)$ \\
$s$ [m/s] & $\mathcal{MJ}\,(1, {\rm RV}_{{\rm max},i}-{\rm RV}_{{\rm min},i})$ \\
$\sigma_{\rm wn}$ [m/s] & $\mathcal{MJ}\,(1, {\rm RV}_{{\rm max},i}-{\rm RV}_{{\rm min},i})$ \\
\end{tabular}
\label{table:t3}
\end{table}

We used the open-source software {\texttt{kima}}\footnotemark\footnotetext{\url{https://github.com/j-faria/kima}} \citep{FSFB2018} to perform Bayesian statistical analysis of all simulated RV datasets. These were analysed assuming the meta-model described in sub-section 2.3, with ${\rm n_p}$ now becoming effectively a label identifying mutually exclusive models. We then have ${\rm n_p}={\rm n_t}+{\rm n_{nt}}$, where ${\rm n_t}$ and ${\rm n_{nt}}$ are, respectively, the number of transiting and non-transiting planets in each system. While the former is fixed, to either $1$ or $2$, we let the latter vary between $0$ and $5$, with equal prior probability assigned to each possible value. This means that we assume a priori all the planets detected in transit to have the status of confirmed planets from the point of view of the RV data analysis. The orbital periods, $P$, and times of mid-transit with respect to the transiting planets were assigned Gaussian priors, centred on the values provided by \citet{Barclay2018}, and with standard deviations of $0.001$ days (which is the typical level of uncertainty expected from TESS data). Knowledge about the time of mid-transit effectively constrains the mean anomaly at some particular time of choice, $M_0$, given the other orbital parameters. For the planets without transit information, the orbital periods were assigned log-uniform (often called Jeffreys) priors between 1 and 10 000 days. For all planets, we assumed modified log-uniform distributions for the RV semi-amplitudes, $K$, and the standard deviations associated with the Gaussian white noise contributions, both $s$ and $\sigma_{\rm wn}$, with the knee located at 1 m/s, while limited above by $1000$ m/s in the case of $K$ and the RV span, i.e. the difference between the RV maximum and minimum, for each RV dataset $i$, ${\rm RV}_{{\rm max},\,i}-{\rm RV}_{{\rm min},\,i}$, in the case of both $s$ and $\sigma_{\rm wn}$. These modified distributions are defined until the lower limit of $0$ m/s. The prior for the orbital eccentricities was set to a half-Gaussian with $\sigma=0.32$ for the transiting planet in systems with only one, and $\sigma=0.083$ for both transiting planets in systems with two, as suggested by \citet{V2019}, and to a Kumaraswamy distribution \citep{Kumaraswamy1980}, with shape parameters $\alpha=0.867$ and $\beta=3.03$, for all the possible extra, non-transiting planets \citep[which is similar to what was proposed in][]{Kipping2013}. Finally, the priors we assumed for the Gaussian Process parameters $\eta_1$, $\eta_2$, $\eta_3$ and $\eta_4$ were, respectively, log-uniform within the interval $[0.1\,{\rm m/s},\,{\rm RV}_{{\rm max},\,i}-{\rm RV}_{{\rm min},\,i}]$, log-uniform between $P_{{\rm rot},i}$ and $5P_{{\rm rot},i}$, Gaussian with mean equal to $P_{{\rm rot},i}$ and standard deviation set to $5\sigma_{{\rm rot},i}$, and log-uniform within the interval $[1/e,\,e]$. Most other parameters are assigned uniform priors between sensible limits, as can be seen in Table \ref{table:t3}.

From the computational point of view, the total amount of datasets is 3 (scheduling strategies ) $\times$ 2 (stellar activity noise models) $\times$ 10 (simulations) $\times$ 50 (systems) = 3000, each containing between 4 and 49 measurements (most being around 22). On a single processor, {\texttt{kima}} requires a few hours to yield converged posterior probability distributions with respect to all model parameters. Thus, it would have been infeasible to perform the analysis sequentially on a single computer. As a result, we adopted a full Cloud architecture by exploiting the services offered by the commercial platform Amazon Web Services (AWS). In particular, since the analysis of each dataset by is independent from the others, we used the architecture described in \citet{Landoni2019} to run parallel applications by using clusters offered by AWS. In this particular case, we used a cluster of 25 instances on AWS, each of them equipped with 64 vCPU and 256 GB of RAM. This allowed the analysis of all 3000 datasets by {\texttt{kima}} in less than 5 hrs and consuming roughly 8000 CPU/hrs in the process. This approach is particularly useful when the full analysis needs to be re-done, for some reason, since it can be performed quickly, while keeping the overall price low.

Our main objective with this work is to compare the different scheduling strategies with respect to: (1) the strength of the expected constraints on the values for the mass and orbital parameters of the planets that are known to transit; (2) the number of detected non-transiting planets, as well as the strength of the expected constraints on the values associated with the respective mass and orbital parameters. These criteria are linked, given that a decision on how many extra planets have been detected is effectively equivalent to choosing the model, with some label ${\rm n_p}$, to be used for parameter estimation. We choose to base such decision on the comparison between the Bayesian evidence or marginal likelihood, i.e. the constant which normalizes the joint posterior distribution, for models with associated consecutive values for ${\rm n_p}$, starting with ${\rm n_p}={\rm n_t}$, i.e. ${\rm n_{nt}}=0$. Because we are assigning equal prior probabilities to all models with respect to the same star, comparing evidences is equivalent to determining the so-called Bayes Factor, $\mathcal{B}$, which is then just equal to their ratio. Its value can be interpreted through the scale introduced by \citet[e.g.][]{Jeffreys1998} (see also \citealt{KR1995}), according to which a Bayes factor of at least 150 between models with associated consecutive values for ${\rm n_p}$ is required in order to claim a planet detection \citep[e.g.][]{FBH2011,FH2013,BD2015}. Note that this procedure never leads to more detections of non-transiting planets than their true number, in any given system. The correspondence between detected and existing non-transiting planets in a system is based on the proximity of values for $K$ and $P$. This criteria never lead to ambiguous cases in our simulations, as a result of the large difference in the values of one or both of these quantities in the few systems with more than one non-transiting planet. 
\begin{figure*}
\includegraphics[width=\linewidth]{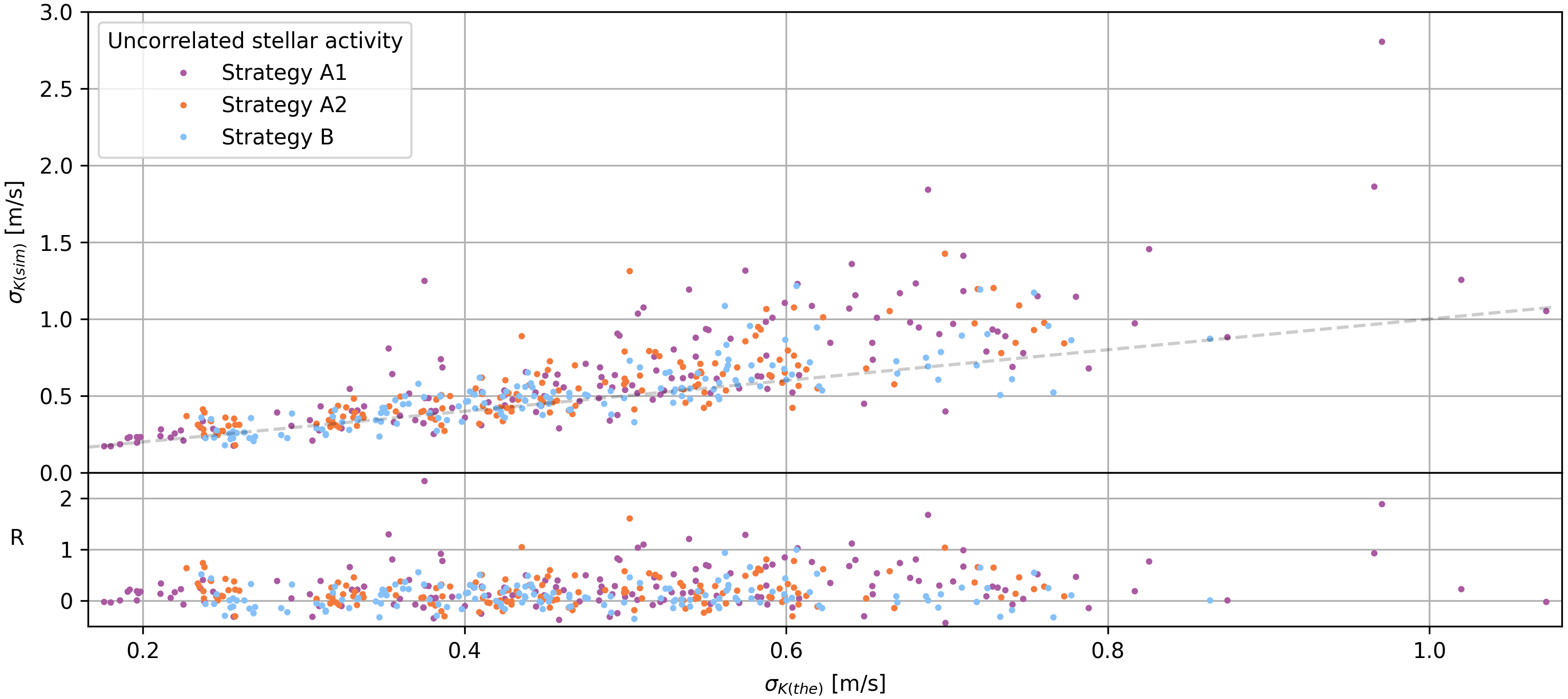}
\includegraphics[width=\linewidth]{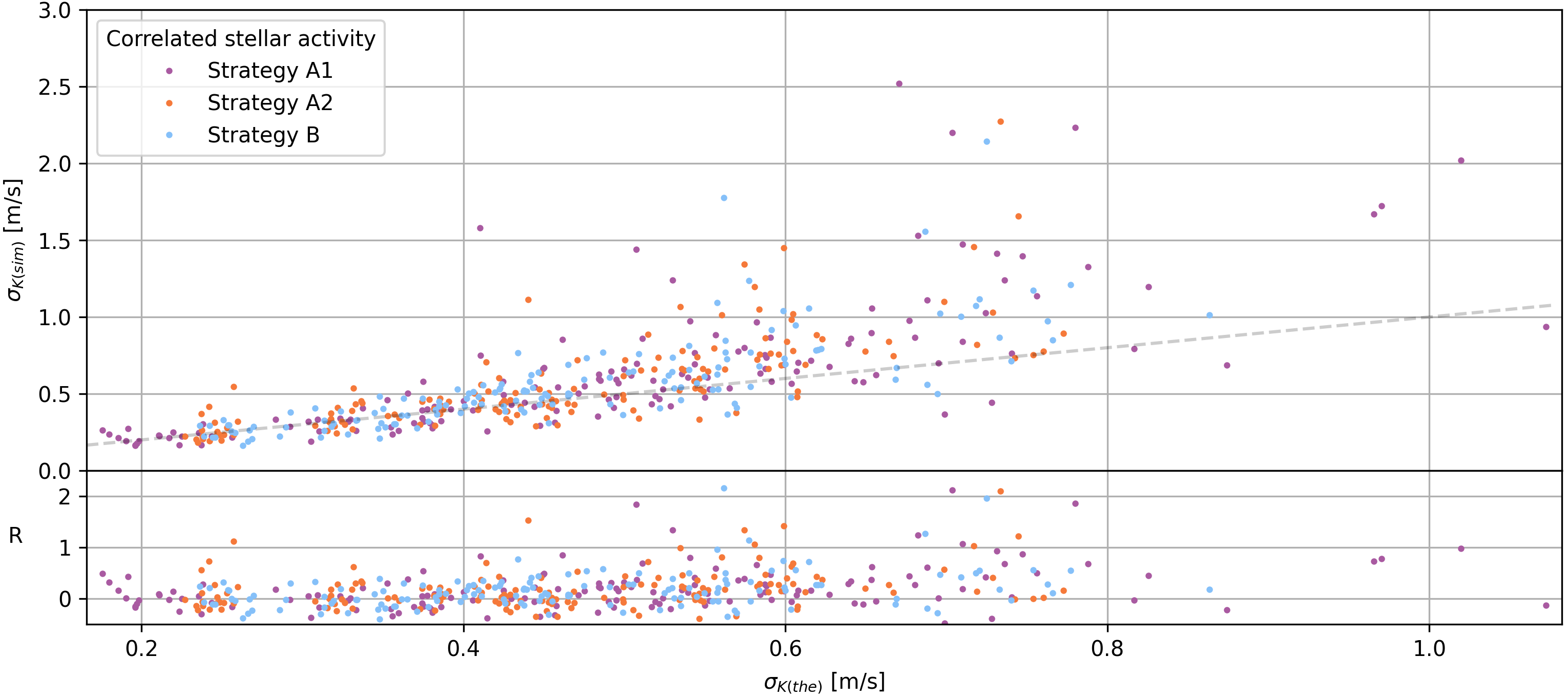}
\caption{In each plot, the upper panel shows the relation between the precision with which $K$ is estimated based on the simulated RV datasets, $\sigma_{K ({\rm sim})}$, and the theoretical precision expected under Eq. \ref{sigcloutier1}, $\sigma_{K ({\rm the})}$, for the transiting planets in the $450$ datasets pertaining to the $15$ systems with only one (transiting) planet. The result of the analysis for each of those datasets is represented by a point, whose colour is associated with the scheduling strategy used: magenta for strategy A1; orange for strategy A2; cyan for strategy B. In each plot, the lower panel shows the residuals $R\equiv[\sigma_{K ({\rm sim})}-\sigma_{K ({\rm the})}]/\sigma_{K ({\rm the})}$. In the top figure, the RV datasets were simulated assuming the stellar activity induced RV variations are uncorrelated, while in the bottom figure those variations were assumed correlated. Equality between ordinate and abscissa is represented by the dashed line. Note that the only difference between both plots is in the values of the ordinate.}
\label{fig:precision_single}
\end{figure*}

\begin{figure*}
\includegraphics[width=\linewidth]{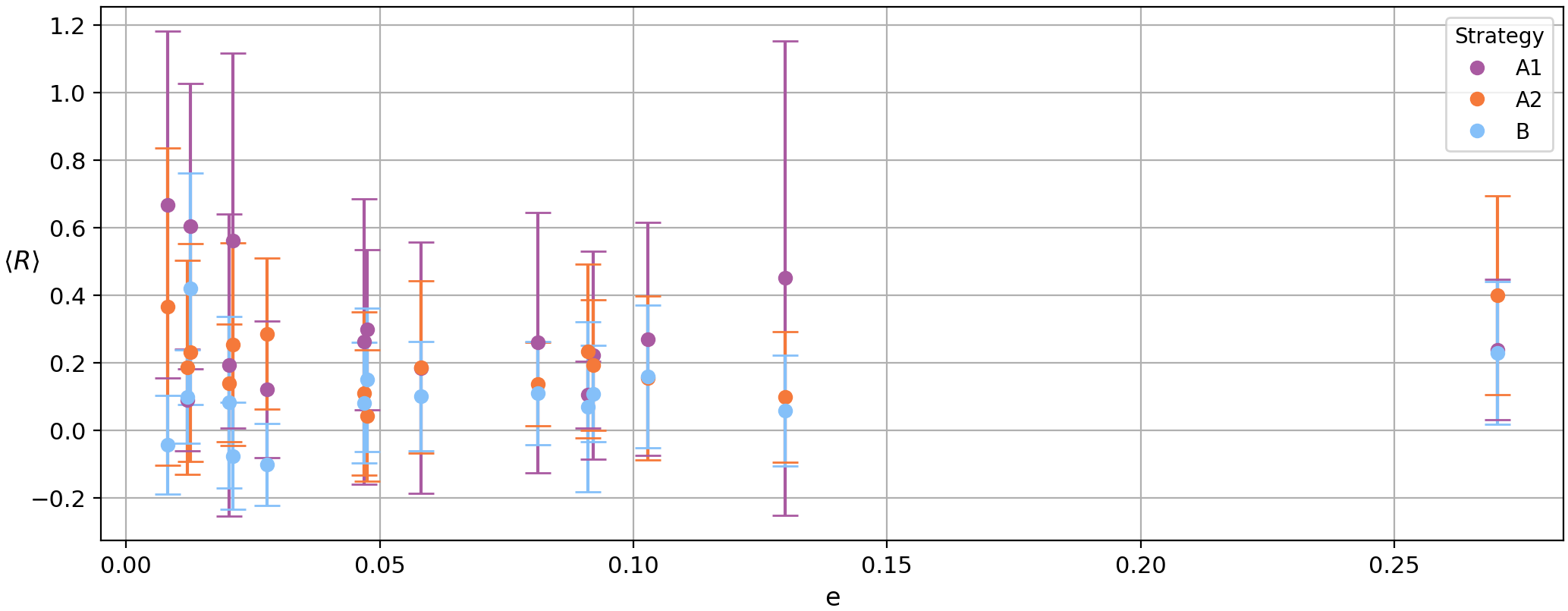}
\caption{Each point represents the relative difference between $\sigma_{K ({\rm sim})}$ and $\sigma_{K ({\rm the})}$, as represented in the lower panel of the top plot in Fig. \ref{fig:precision_single}, averaged over the 10 RV datasets simulated for each system per scheduling strategy (colour-coded as in Fig. \ref{fig:precision_single}), as a function of the eccentricity of the transiting planet in the system. }
\label{fig:precision_single_wn_ecc}
\end{figure*}

It should be noted that by using the full RV datasets in the analysis we are effectively assuming that there is neither partial or full loss of planned RV measurements due to adverse weather conditions or technical problems. Here partial also means substantial degradation of the expected RV measurement uncertainty due to photon-noise, as a result of very bad seeing ($>1.3''$) or thick cirrus clouds. Although such assumption is unrealistic, the loss should not amount to more than 10 percent of the expected data, according to the ESO annual reporting on the operational conditions at Paranal\footnotemark\footnotetext{\url{https://www.eso.org/public/products/annualreports/}}. Therefore, on average this should affect only a couple of RV measurements per target in three years. In any case, which RV measurements are affected or lost, as a result of such effects, will not be correlated with the actual scheduling strategy chosen to be implemented. Therefore, the data loss will impact in a similar way the information about planetary masses and orbital parameters that can be recovered under each scheduling strategy. Thus, we decided to ignore it in our analysis since our interest is on the comparison of the relative merit of different scheduling strategies. However, in a practical context, this issue can be addressed and its impact minimised by rescheduling all future observations after some amount of the planned RV measurements are performed, and taking into account which were not or badly affected.

\begin{figure*}
\includegraphics[width=\linewidth]{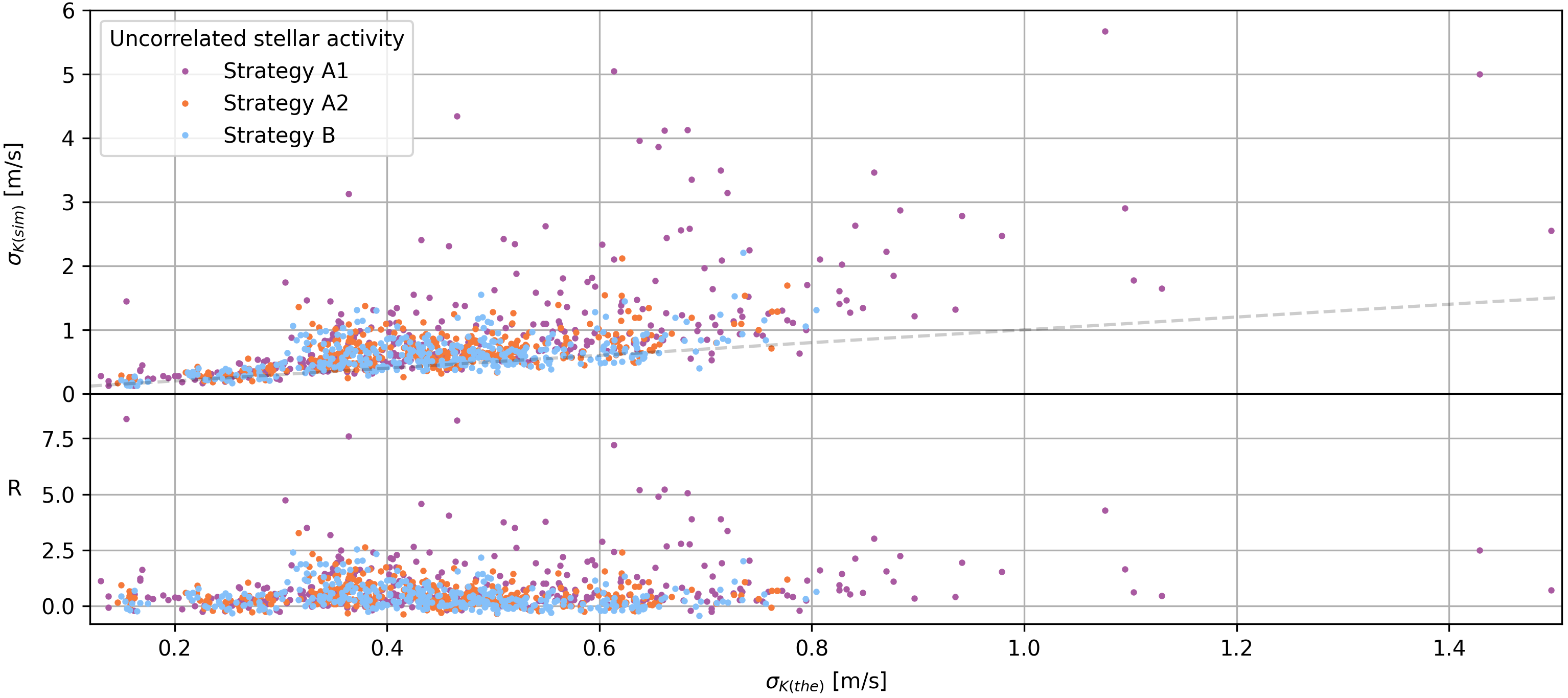}
\includegraphics[width=\linewidth]{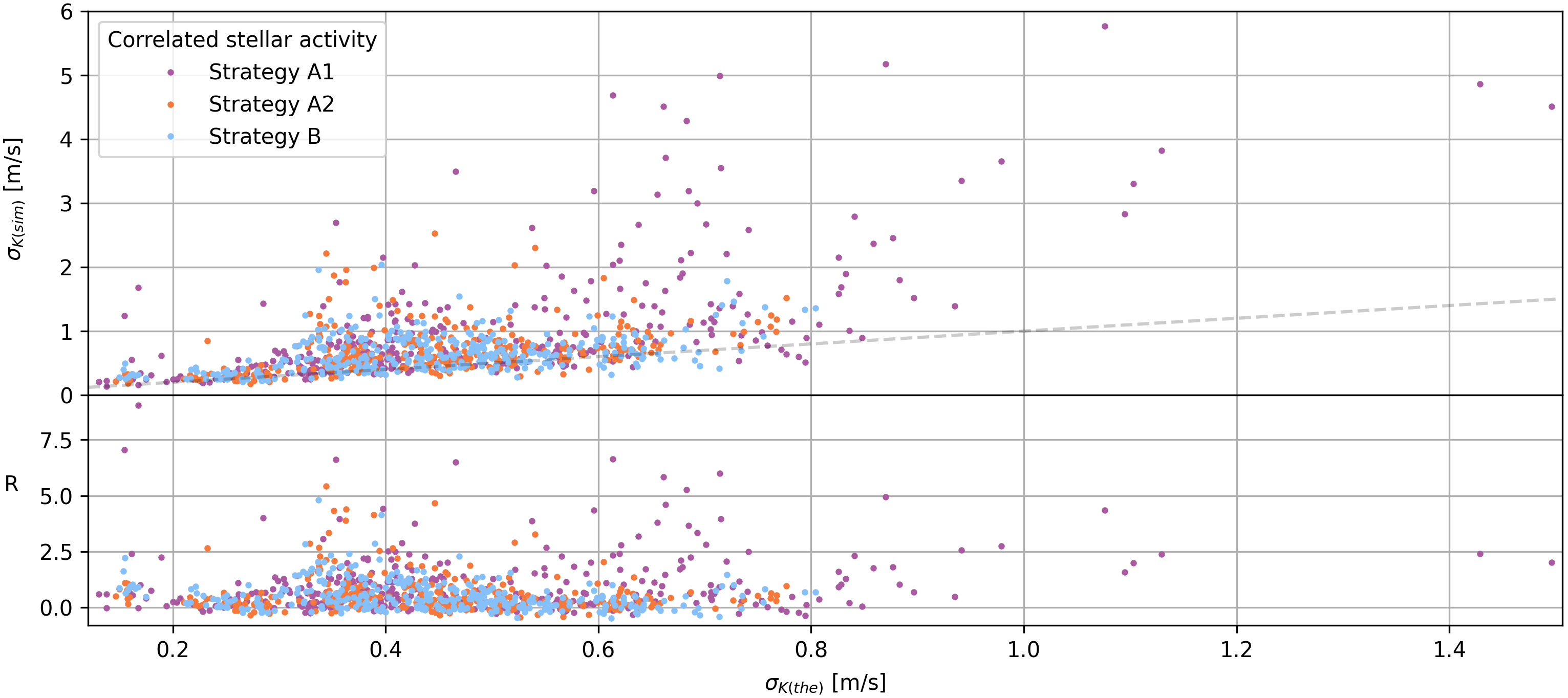}
\caption{The top and bottom plots show the same information as in Fig. \ref{fig:precision_single}, but now each point refers to the results for the $38$ transiting planets in the $35$ systems with more than one planet. Again, note that the only difference between both plots is in the values of the ordinate.}
\label{fig:precision_multi}
\end{figure*}

\subsection{Comparison with theoretical expectations}

\begin{figure*}
\includegraphics[width=\linewidth]{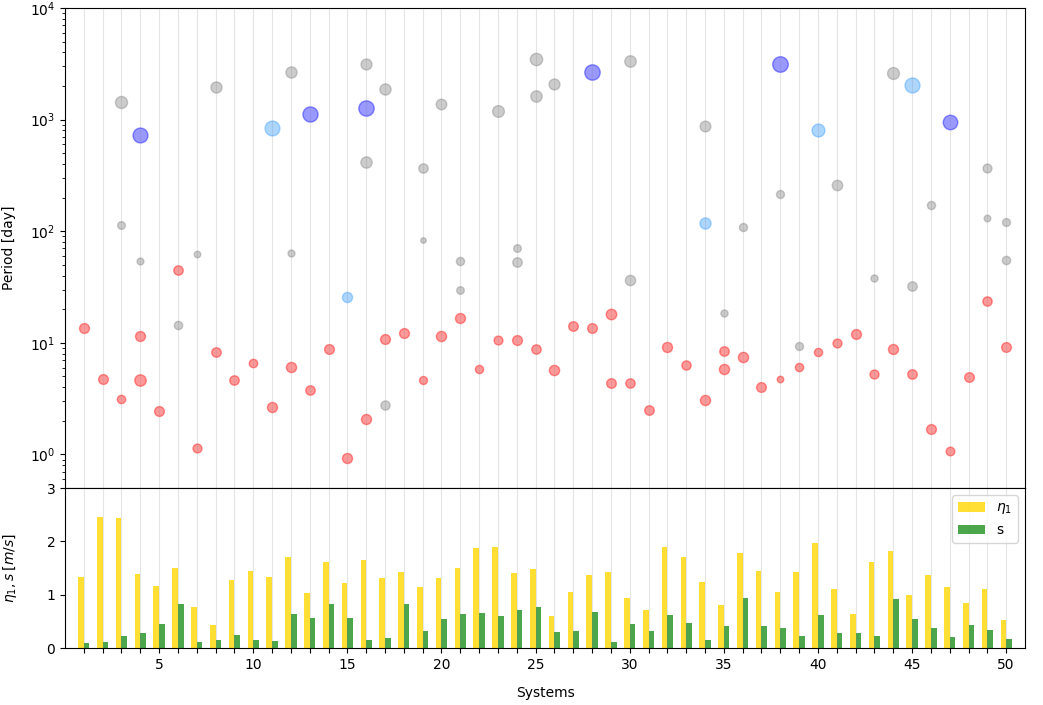}
\caption{The upper panel shows the planetary system architecture as a function of orbital period. Light red circles represent transiting planets, while non-transiting planets are represented by grey circles, when never detected, light blue circles, when detected at least once, and violet circles, when always detected, with respect to the $3\times10$ simulations carried out for the three scheduling strategies. The planets coloured light blue were detected with strategies [A1, A2, B], respectively, the following number of times: system 11 [7, 10, 10]; system 15 [1, 0, 0]; system 34 [0, 0, 1]; system 40 [8, 5, 7]; system 45 [6, 10, 10]. The size of each circle is proportional to the mass of the respective planet. In the lower panel, the values for the correlated, $\eta_1$ (yellow), and non-correlated, $s$ (green), components of the stellar activity induced RV variations are represented for the stars in our sample. Each star and its associated planetary system is identified by an incremental number where 1 corresponds to the lowest TESS ID number and 50 to the highest TESS ID number in our sample.}
\label{fig:d_gp}
\end{figure*}

The RV variations induced by a transiting planet with known orbital period, $P$, and time of mid-transit, $T_{\rm transit}$, depend only on the (unknown) value of $K$ if the orbit is assumed circular, 
\begin{equation}
v_{\rm r}(t_{\rm i})=K\sin\phi(t_{\rm i},P,T_{\rm transit})\,,
\label{kepleriancircular}
\end{equation}
with the true anomaly at time $t_{\rm i}$ given by
\begin{equation}
\phi(t_{\rm i})=2\pi(t_{\rm i}-T_{\rm transit})/P\,.
\label{trueanomalycircular}
\end{equation}
Further assuming that the measurements of such RV variations are affected by uncertainties that are independent and identically Gaussian distributed, i.e. Gaussian white noise, it can then be shown \citep[e.g.][]{Cloutier2018} that the theoretically expected (a posteriori) absolute precision in the estimation of $K$, which we will call $\sigma_{K{\rm (the)}}$, is given by 
\begin{equation}
\sigma_{K{\rm (the)}}=\left\{\sum\limits_{\rm i=1}^{N_{\rm RV}}\left\{\frac{\sin[\phi(t_{\rm i},P,M_0)]}{\sigma_{\rm eff}(t_{\rm i})}\right\}^{2}\right\}^{-1/2}
\label{sigcloutier1}
\end{equation}
where $\sigma_{\rm eff}(t_{\rm i})$ is the effective measurement uncertainty with respect to the planet-induced RV at time $t_{\rm i}$, and $N_{\rm RV}$ is the number of RV measurements considered. The former can result from several Gaussian white noise contributions, and is just the standard deviation associated with the Normal distribution that describes the full uncertainty with respect to the RVs induced just by the transiting planet. 

Assuming the effective RV measurement uncertainty is approximately constant with time, the minimization of $\sigma_K$ demands sampling the orbital phase space only when the radial velocity reaches its maximum absolute value. This corresponds to what is usually known as quadrature sampling, whereby RV measurements are performed only when the transiting planet is at right-angles with respect to our line-of-sight to the star. In this case,
\begin{equation}
\sigma_{K{\rm (the)}}=\frac{\sigma_{\rm eff}}{\sqrt{N_{\rm RV}}}
\label{sigcloutier2}
\end{equation}
which is what one would expect under the central limit theorem if each RV measurement corresponds in fact to a direct estimation of $K$. Thus, quadrature sampling is the optimal procedure if (1) $\sigma_{\rm eff}$ is independent of the measurement time and (2) all the conditions under which Eq. \ref{sigcloutier1} was derived are assured. If the former is not true, but the later is, then some RV measurements off quadrature can actually yield more information about $K$, if the associated RV measurement uncertainty is sufficiently smaller. If condition (2) is not true, quadrature sampling can lead to biased results. This can be easily seen in the case of non-circular orbits. If these are only sampled in quadrature, the estimates for the eccentricity $e$ and $K$ will be significantly degenerate, reaching complete degeneracy when the argument of periastron is such that the orbital semi-major axis becomes aligned with our line-of-sight to the star. These degeneracies allow very high values simultaneously for $e$ and $K$, given the RV quadrature data. Although such combinations could be disfavoured a priori, reducing their impact in a posteriori estimates of $e$ and $K$, like their means, medians or modes, they will nevertheless tend to bias high any such estimates.

If the orbital phase-space is sampled uniformly, while still assuming the effective RV measurement uncertainty to be approximately constant with time, Eq. \ref{sigcloutier1} then implies a decrease by a factor of $\sqrt{2}$ in the absolute precision with which $K$ can be estimated, i.e.
\begin{equation}
\sigma_{K{\rm (the)}}=\sigma_{\rm eff}\sqrt{\frac{2}{N_{\rm RV}}}\,,
\label{sigcloutier3}
\end{equation}
with respect to quadrature sampling. This decrease can be understood by realizing that now a significant fraction of the RV measurements are made close to anti-quadrature, when star and planet are aligned along the line-of-sight, and thus the RV signal-to-noise ratio, i.e. that between the expected RV amplitude and $\sigma_{\rm eff}$, becomes much smaller with respect to RV measurements made in quadrature. Therefore, less information about $K$ will be gathered, on average, per RV measurement.

However, in the case of non-circular orbits, uniform-in-phase sampling partially lifts the degeneracies between argument of periastron, $e$ and $K$, so much so the denser the sampling. Thus, this type of sampling does not lead to such strong biases as quadrature sampling. We can then conclude that if there is some significant probability of $e$ being different from zero, one should opt for uniform-in-phase rather than quadrature sampling to ensure that estimates of $K$ in particular are as unbiased as possible. Nevertheless, uniform-in-phase sampling is still not optimal if one wants to maximize the amount of information that can be gathered through RV measurements about the orbital parameters, when in the presence of non-circular orbits. This is again, in part, the result of variations in the RV signal-to-noise ratio across the orbital phase-space. The optimal sampling solution could be found using the tools of bayesian experimental design \citep[e.g.][]{Ford2008,Loredo2012,Hees2019}. In the absence of any RV information it would depend critically on the assumed prior distributions for the orbital parameters. But as each RV measurement is obtained, the optimal sampling solution can be continuously updated, converging to the same solution whatever the assumed prior distributions. Unfortunately, bayesian experimental design carries a very high computational cost, which is why we did not consider using it in this work.

Our results present an unique opportunity to test the impact of different effects on the absolute precision with which $K$ can be estimated for transiting planets, $\sigma_K$. We start by considering the results of the analysis of the $3\times15\times10=450$ RV datasets pertaining to systems which contain only one (transiting) planet and for which it was assumed just Gaussian white noise. In the the top plot in Fig. \ref{fig:precision_single}, the results of each of those 450 analysis is represented by a point. The ordinates correspond to the values of $\sigma_K$ that result from the analysis of the datasets, which we will denote as $\sigma_{K{\rm (sim)}}$, while the abscissas are calculated using Eq. \ref{sigcloutier1} given the characteristics of each dataset. The later would correspond to the expected value of the former if the orbits were circular. If this was the case, we should see only stochastic variations about zero for the deviations of the ordinates with respect to the abscissas. Although the lower panel of the top plot in Fig. \ref{fig:precision_single} seems to suggest otherwise, in particular for the datasets acquired under strategy A1, the variations seen are indeed statistically compatible with the expected value for $\sigma_{K{\rm (sim)}}$ being well approximated by Eq. \ref{sigcloutier1}, even though the orbits of the planets considered are not circular: for strategies A1, A2 and B, the mean and standard deviation of the residuals shown in the lower panel of the top plot in Fig. \ref{fig:precision_single} are $0.30\pm0.36$, $0.20\pm0.25$ and $0.10\pm0.19$. Nevertheless, these values suggest that how well Eq. \ref{sigcloutier1} predicts $\sigma_K$ depends on how close the sampling of the phase-curve is to uniform. It also seems to depend on how high is the information content of the RV measurements, given the decreasing scatter in the residuals as the expected value for $\sigma_{K{\rm (sim)}}$  decreases. All these conclusions seem to be true irrespective of the eccentricity of the planets considered, given that this does not seem to be correlated with the magnitude of the residuals, as can be seen in Fig. \ref{fig:precision_single_wn_ecc}.

The results of the analysis of the RV datasets pertaining to the $35$ systems which contain more than one planet, and for which it was assumed just Gaussian white noise, allow for the characterization of the impact of extra planets in the absolute precision with each $K$ can be recovered for the $38$ transiting planets in those systems. In the upper panel of the top plot in Fig. \ref{fig:precision_multi}, the results associated with each of those $3\times38\times10=1140$ analysis are represented by points, with the coordinates having the same meaning as in Fig. \ref{fig:precision_single}. It is perceptible an increase in scatter, as well as a higher systematic (positive) difference between the value for $\sigma_K$ that results from each analysis and the value given by Eq. \ref{sigcloutier1}. Now, the mean and standard deviation of the residuals shown in the lower panel of the top plot in Fig. \ref{fig:precision_multi} are $0.95\pm0.82$, $0.53\pm0.38$ and $0.42\pm0.36$, for strategies A1, A2 and B, respectively. Again, there seems to be no correlation between the magnitude of the residuals and the eccentricity of the planets considered, all below $0.3$ and with a mean value of $0.08$, very similar to what is obtained ($0.07$) for the $15$ lone transiting planets.

Finally, in the bottom plots of Figs. \ref{fig:precision_single} and \ref{fig:precision_multi} we show the results for the same two sets of planets, but when stellar activity induced RV variations are assumed correlated and jointly modelled as a GP. The mean and standard deviation of the residuals shown in the bottom panels are, respectively, for strategies A1, A2 and B:  $0.23\pm0.39$, $0.18\pm0.29$ and $0.16\pm0.24$, for the $15$ lone transiting planets; $0.98\pm0.87$, $0.58\pm0.53$ and $0.53\pm0.40$, for the $38$ transiting planets with companions. Remarkably, these numbers are very similar to those obtained when the stellar activity induced RV variations are assumed non-correlated and modelled as Gaussian white noise. This is just a reflection of the fact that the abscissas are the same, and the ordinates are very similar. As we will later discuss, when $\sigma_K$ is averaged over all transiting planets, it differs by $0.03$ at most (less than $10\%$), for any of the three scheduling strategies, between what is obtained under the two contrasting assumptions about the characteristics of the stellar activity induced RV variations. This indicates that, as long as these variations are correctly modelled (which is difficult to ascertain for any particular star besides the Sun), and there is enough information in the RV measurements (as seem to be the case in our simulated datasets), their impact on the absolute precision with which $K$ can be recovered for transiting planets is essentially independent of whether they are correlated or not (but dependent on their amplitude).

In summary, we find that Eq. \ref{sigcloutier1} yields a good approximation to the expected absolute precision with which $K$ can be recovered for transiting planets, even in the presence of mild eccentricity (less than $0.3$) and realistic correlated RV variations due to stellar activity. Nevertheless, large deviations (typically up to a $50\%$ increase) from the expectation are possible, the more so the less uniform is the sampling of the orbital phase curves and the smaller the information content of the RV measurements. As expected, the presence of extra planets in a system leads to an increase both in the magnitude and scatter of $\sigma_K$ with respect to the value expected under Eq. \ref{sigcloutier1}. For the type of planetary systems considered, this increase is reflected in a typical underestimation of $\sigma_K$ for the transiting planets, between $40\%$ and $100\%$, and similar scatter, with higher values corresponding to orbital phase curves sampled less uniformly and less informative RVs. In any case, we would like to stress that these conclusions are conditional on the assumption that the RV data generating mechanisms, in particular those associated with stellar activity and the instrumentation used, are well approximated by the assumed model.

\subsection{Results obtained assuming stellar activity correlated noise}

We will now focus the discussion on the results obtained when the RV variations induced by stellar activity were assumed to be correlated, generated by a Gaussian Process with non-zero covariance terms, given that this constitutes the most realistic scenario. In Appendix A, we present and compare with these, the results obtained when stellar activity induced RV variations were assumed to be uncorrelated, akin to Gaussian white noise.

In Figure \ref{fig:d_gp} we show the architecture of the 50 planetary systems we consider. We identify which planets transit, and differentiate between the non-transiting planets that are never detected, sometimes detected or always detected, across all simulations and for all three scheduling strategies. Averaging over the 10 simulations per strategy, a total of $8.2\pm0.6$, $8.5\pm0.5$ and $8.8\pm0.5$ non-transiting planets are detected using strategies A1, A2 and B, respectively, out of the 50 that we simulated orbiting our sample of stars. The numbers provided represent means and standard deviations, and are very similar. The differences are not significant given the variation seen across the simulations.

In order to compare further the results, we define the following quantities, with respect to some planet characteristic $X$, and to a given simulation:

\begin{itemize}
\item absolute bias, $\mathbf{E}[X]-X_{true}$
\item relative bias, $(\mathbf{E}[X]-X_{true})/X_{true}$
\item absolute accuracy, $\mid\mathbf{E}[X]-X_{true}\mid$
\item relative accuracy, $\mid\mathbf{E}[X]-X_{true}\mid/X_{true}$
\item absolute precision, $\sigma_X$
\item relative precision, $\sigma_X/\mathbf{E}[X]$
\end{itemize}
where $X_{true}$, $\mathbf{E}[X]$ and $\sigma_X$ represent, respectively, the true, expected value and standard deviation of $X$. The latter two are estimated given all values for $X$ present in the MCMC output from the {\texttt{kima}} analysis of the dataset associated with the simulation being considered.

\begin{figure}
\includegraphics[width=\columnwidth]{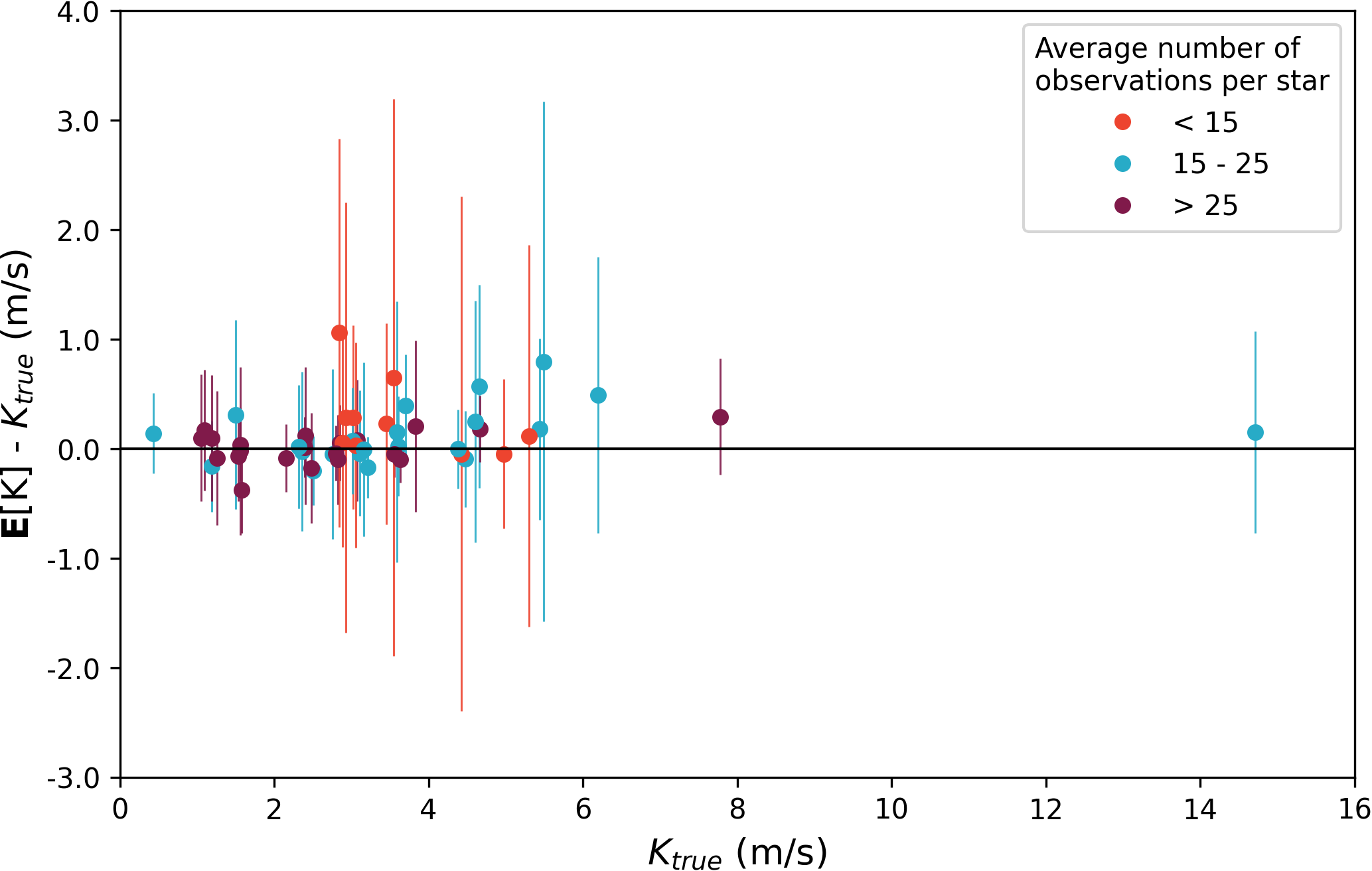}\vspace{0.2cm}
\includegraphics[width=\columnwidth]{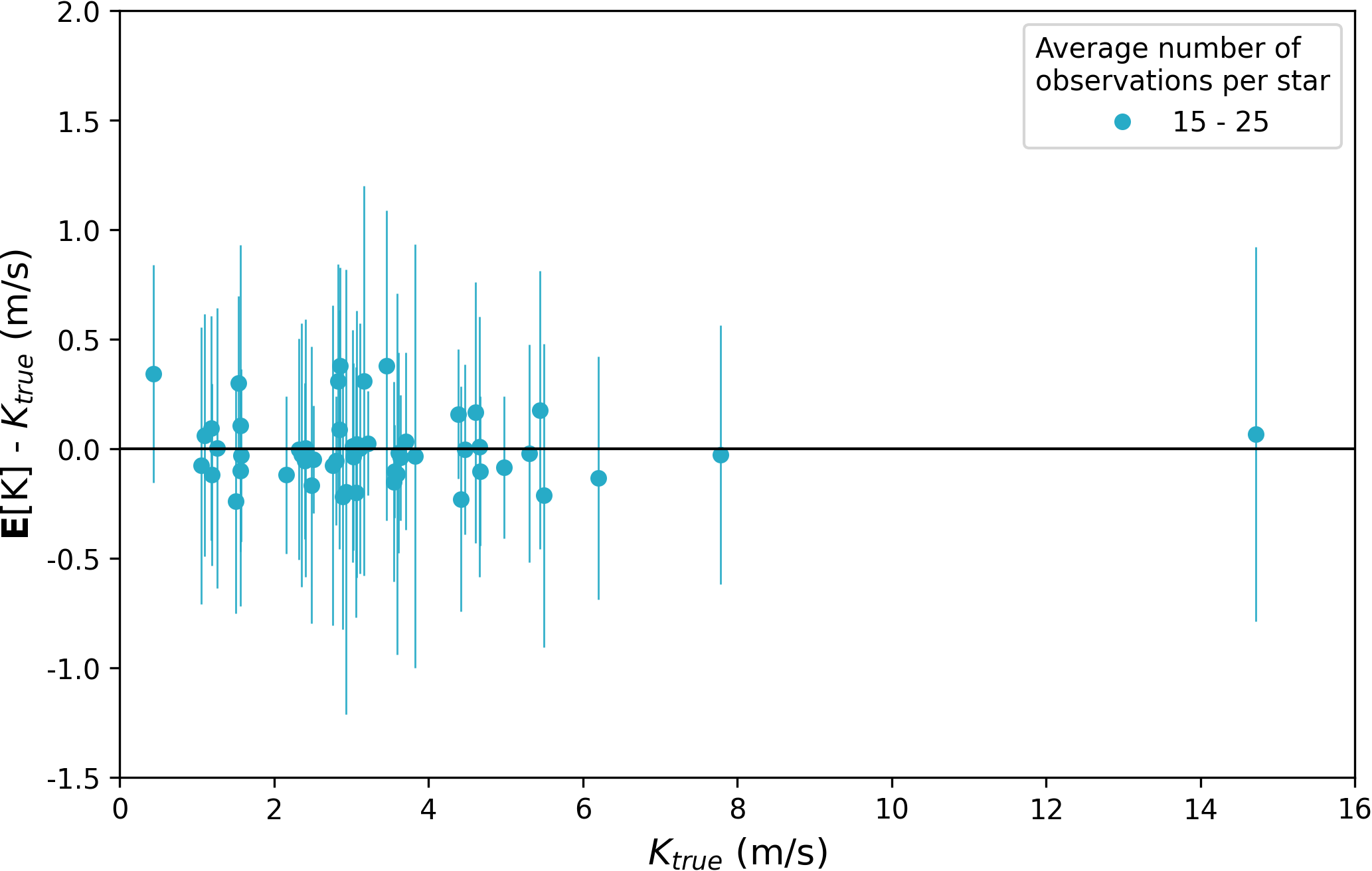}\vspace{0.2cm}
\includegraphics[width=\columnwidth]{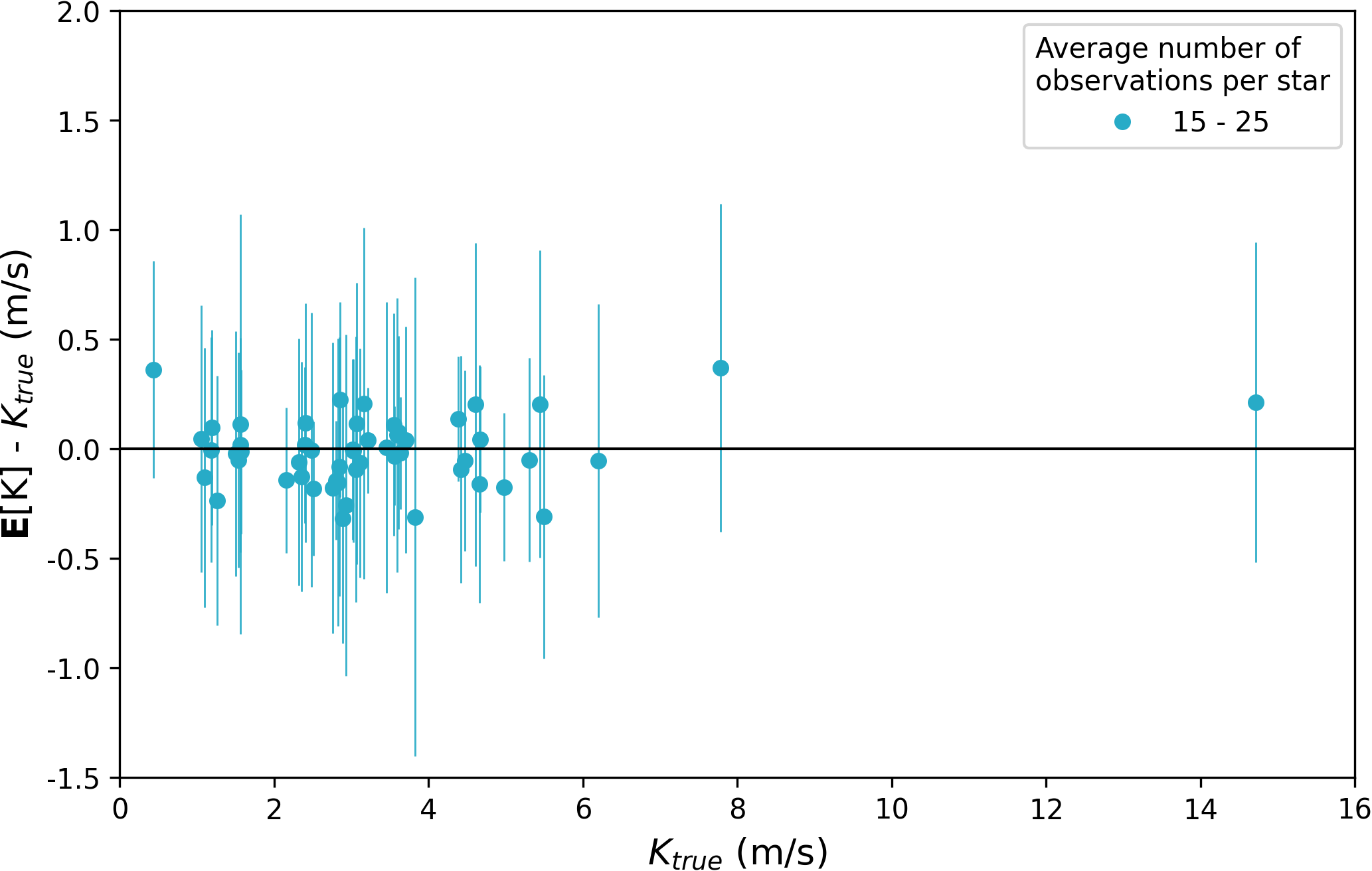}
\caption{Absolute bias, i.e. the difference between the marginal posterior mean and the true value, as a function of the later, for the RV semi-amplitude, $K$, and with respect to the transiting planets. Results averaged over 10 simulations are shown, with the associated standard deviation, for the three scheduling strategies, A1 (upper panel), A2 (middle panel ) and B (lower panel). Colour indicates the number of RV measurements per host star, averaged over the 10 simulations: red, less than 15; blue, between 15 and 25; purple, more than 25.}
\label{fig:K}
\end{figure}

\begin{figure}
\includegraphics[width=\columnwidth]{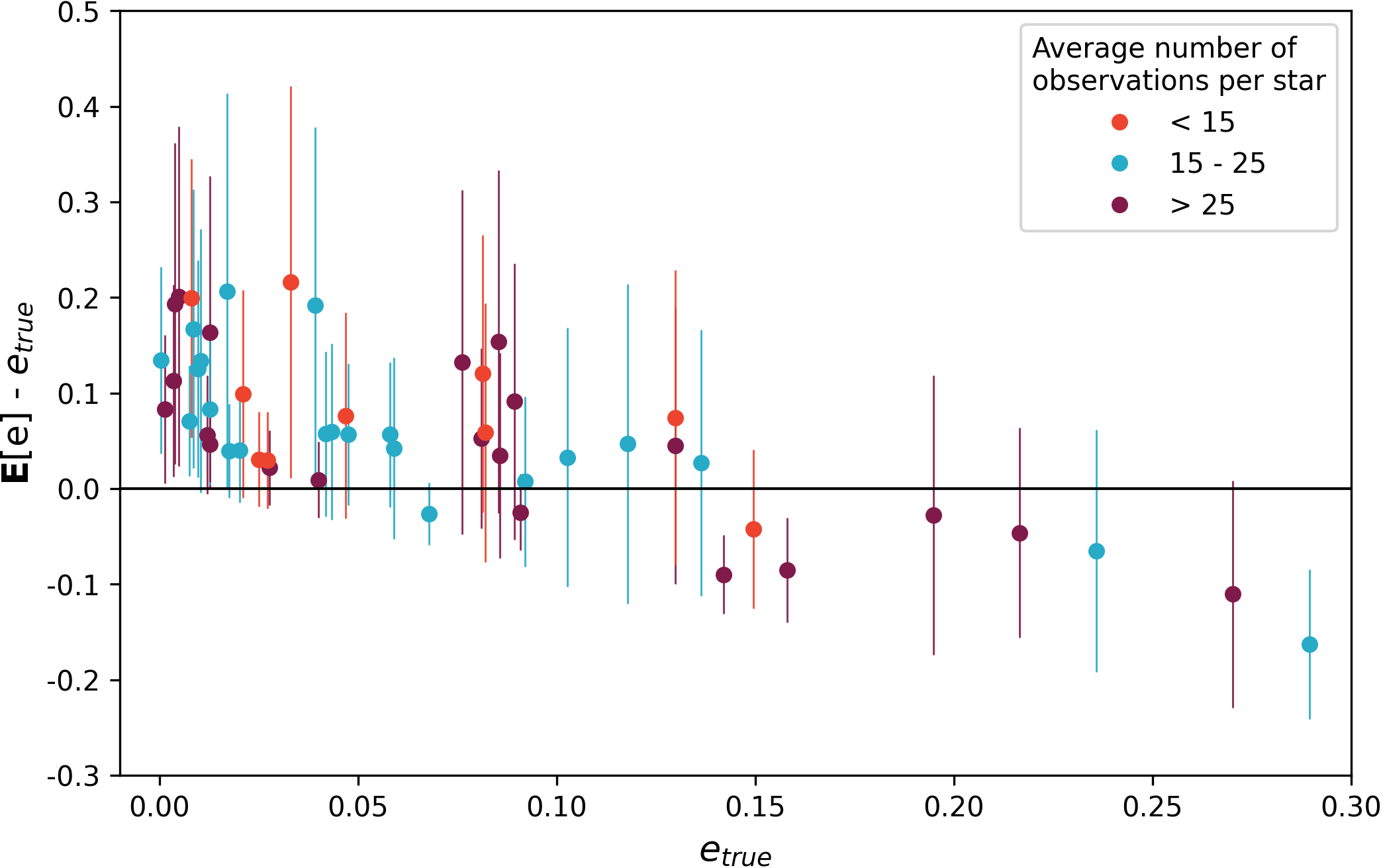}\vspace{0.2cm}
\includegraphics[width=\columnwidth]{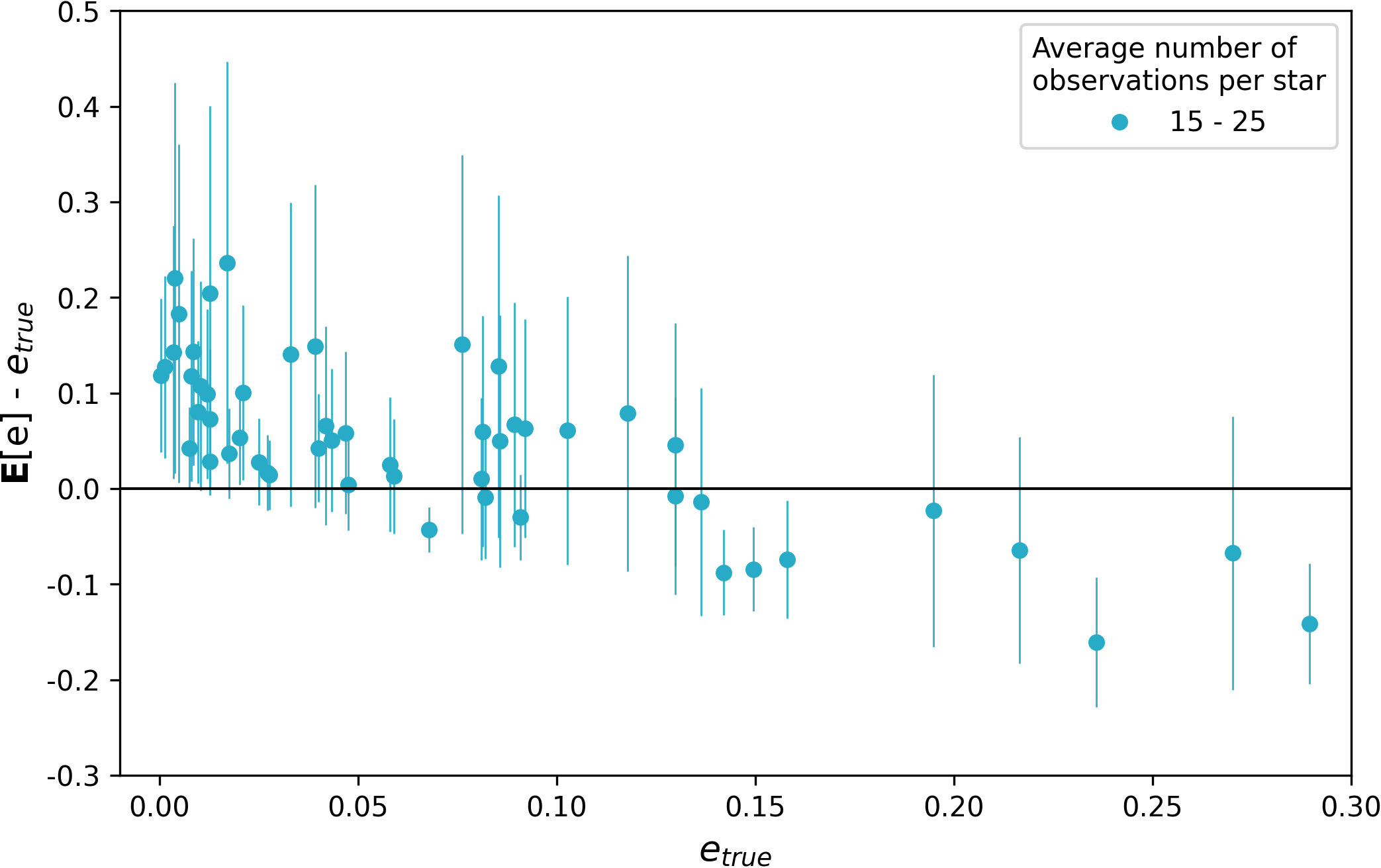}\vspace{0.2cm}
\includegraphics[width=\columnwidth]{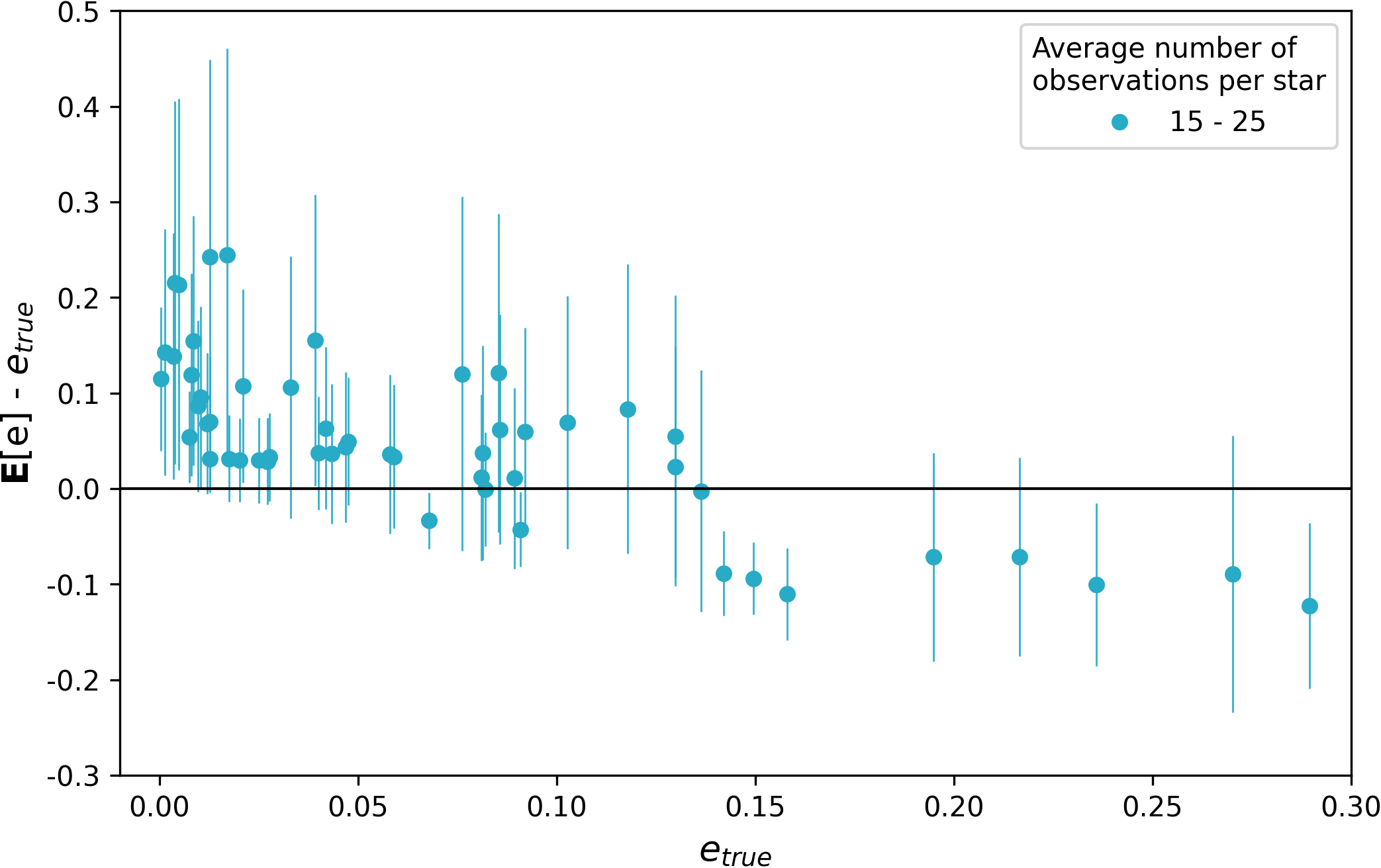}
\caption{Absolute bias, i.e. the difference between the marginal posterior mean and the true value, as a function of the later, for the orbital eccentricity, $e$, and with respect to the transiting planets. Results averaged over 10 simulations are shown, with the associated standard deviation, for the three scheduling strategies, A1 (upper panel), A2 (middle panel ) and B (lower panel). The colour code is the same as in Fig. \ref{fig:K}.}
\label{fig:e}
\end{figure}

\begin{figure}
\includegraphics[width=\columnwidth]{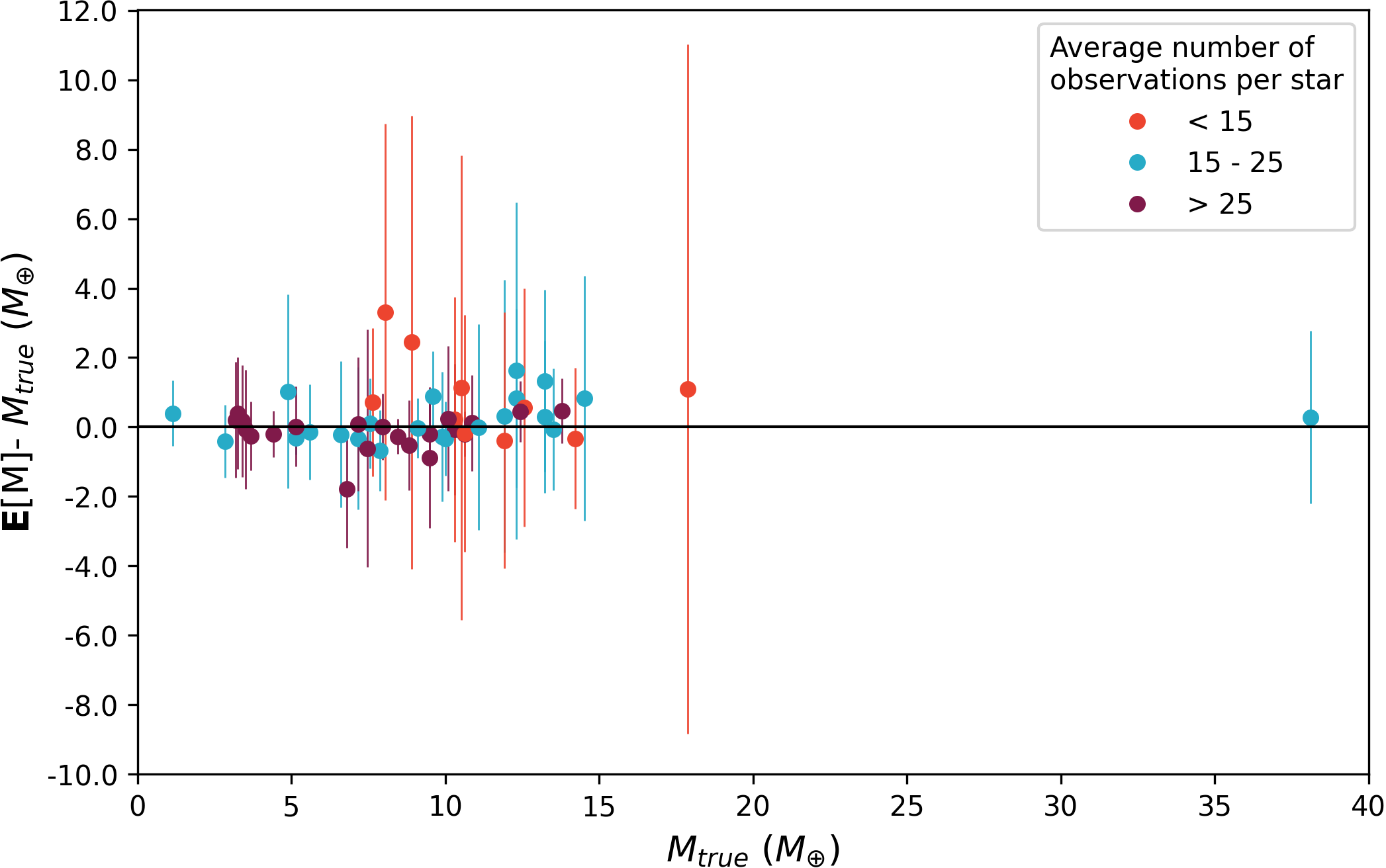}\\
\includegraphics[width=\columnwidth]{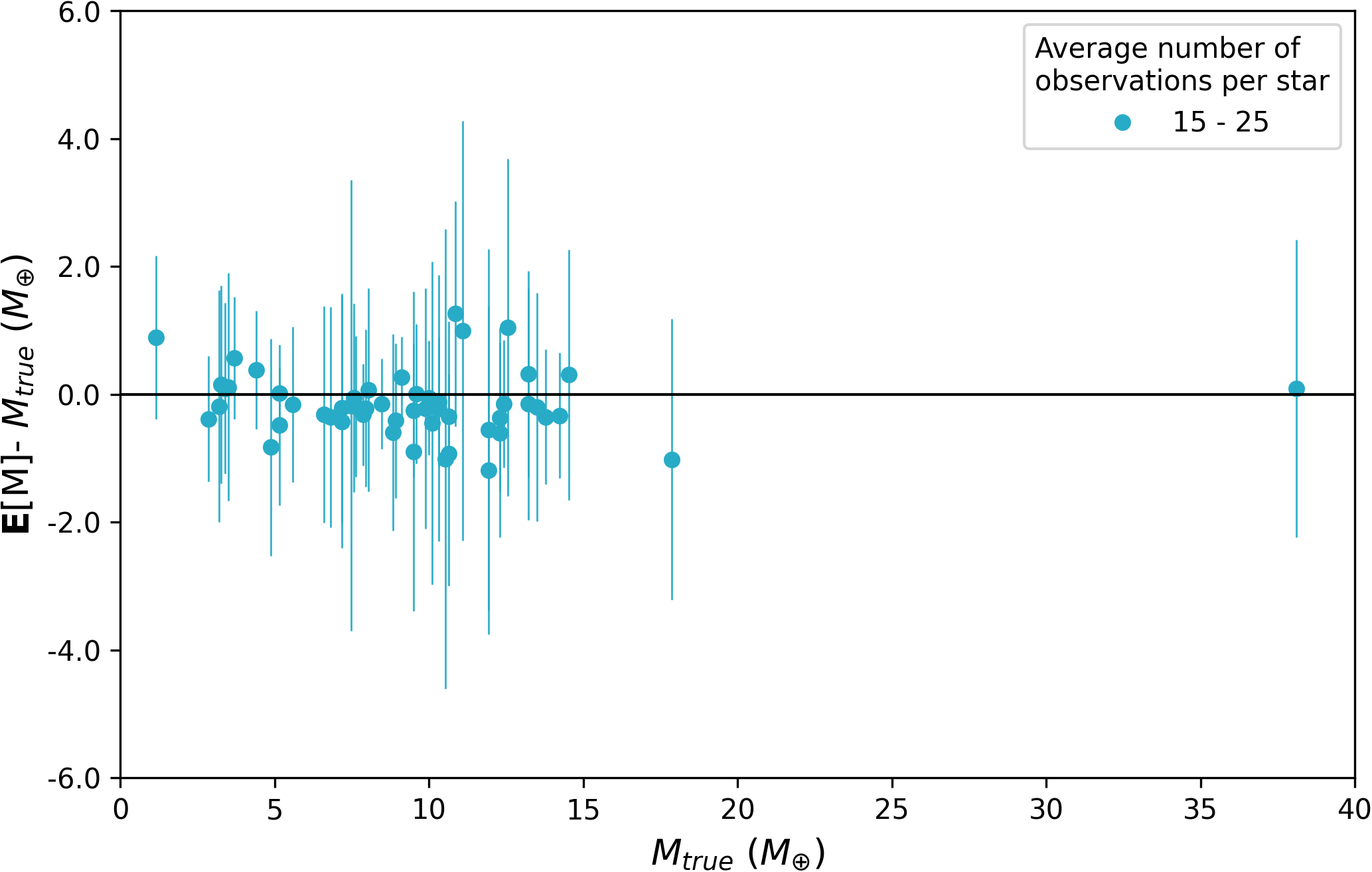}\\
\includegraphics[width=\columnwidth]{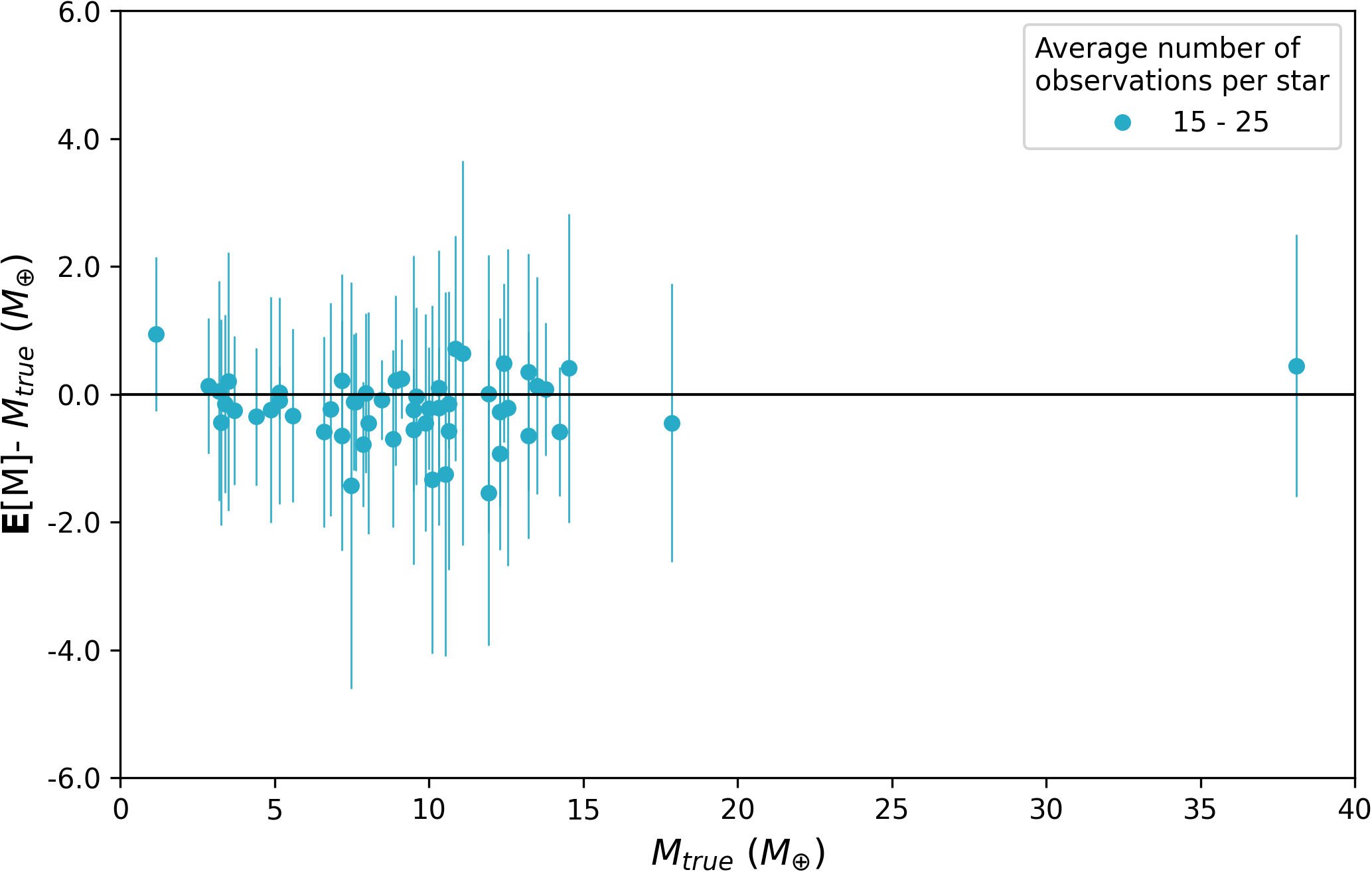}
\caption{Absolute bias, i.e. the difference between the marginal posterior mean and the true value, as a function of the later, for the mass, $M$, and with respect to the transiting planets. Results averaged over 10 simulations are shown, with the associated standard deviation, for the three scheduling strategies, A1 (upper panel), A2 (middle panel ) and B (lower panel). The colour code is the same as in Fig. \ref{fig:K}.}
\label{fig:M}
\end{figure}

In Figures \ref{fig:K}, \ref{fig:e} and \ref{fig:M} we show the means and standard deviations for the distributions of absolute bias associated with the orbital parameters $K$ and $e$, as well as mass, $M$, for the planets that are known to transit and the three scheduling strategies. The averaging is performed over the 10 simulations per system and strategy. Because the absolute bias differs from the expected value by a constant, their distributions have the same standard deviations. Expected values are more scattered (with respect to the true values) and uncertain for some planets in the case of strategy A1 mostly as a result of the respective host stars being systematically under-observed (and others over-observed) with respect to average, due to having shorter (longer) visibility windows. The marginal posteriors used for this exercise, and those that follow, are those associated with the model chosen using the detection procedure for the non-transiting planets previously described, given the result of the Bayesian analysis of each simulated dataset. 

In the upper panel of Table \ref{table:t4}, the absolute and relative bias, accuracy and precision with which $K$, $e$ and $M$ are recovered, averaged over all transiting planets and simulations, is shown for the three strategies. The uncertainties provided are standard deviations, and characterise the dispersion of such values taking into account all transiting planets. They should not be confused with the uncertainties associated with the estimates of $K$, $e$ and $M$ for individual planets, and thus should not be used to draw any conclusions regarding confidence or credible intervals for those quantities. This is particularly true in the case of quantities whose marginal posterior distributions are heavily skewed, like the eccentricity. The same quantities shown in the upper panel of Table \ref{table:t4} are provided in the lower panel, including with respect to the orbital period, $P$, but now averaged over the detected non-transiting planets. In Table \ref{table:t5}, the absolute and relative bias, accuracy and precision with which $\eta_1$, $\eta_2$, $\eta_3$, $\eta_4$ and $s$ are recovered, averaged over all simulations, is shown for the three strategies.

\begin{table*}
\caption{In the upper panel it is shown the absolute and relative bias, accuracy and precision with which $K$, $e$ and mass, $M$, are recovered, averaged over all transiting planets and simulations, for the three strategies. The uncertainties provided are standard deviations, and characterise the dispersion of such values taking into account all transiting planets. The same quantities, as well as the orbital period, $P$, are provided in the lower panel with respect to all detected non-transiting planets. The absolute quantities with respect to $K$, $M$ and $P$ are in units of m/s, $M_{\odot}$ and days, respectively.}
\centering
\begin{tabular}{c c c c c c c c}
\hline\hline
\rule{0pt}{3ex} 
Strategy&Parameter&&Absolute&&&Relative&\\
&&\multicolumn{3}{c}{\rule{0.3cm}{0.0cm}\rule{5.8cm}{0.03cm}}&\multicolumn{3}{c}{\rule{0.3cm}{0.0cm}\rule{5.2cm}{0.03cm}}\\
&&Bias&Accuracy&Precision&Bias&Accuracy&Precision\\
\rule{0pt}{3ex} 
$A_1$ & $K$ & $0.18\pm0.32$ & $0.56\pm0.32$ & $0.91\pm0.60$ & $0.06\pm0.13$ & $0.20\pm0.14$ & $0.33\pm0.24$\\
& $e$ & $0.08\pm0.09$ & $0.11\pm0.06$ & $0.11\pm0.05$ & $15.38\pm55.66$ & $15.52\pm55.62$ & $0.77\pm0.05$\\
& $M$ & $0.20\pm0.79$ & $1.55\pm0.96$ & $2.30\pm1.75$ & $0.02\pm0.11$ & $0.19\pm0.12$ & $0.28\pm0.16$\\
$A_2$ & $K$ & $0.03\pm0.16$ & $0.40\pm0.14$ & $0.64\pm0.24$ & $0.03\pm0.17$ & $0.17\pm0.16$ & $0.28\pm0.25$\\
& $e$ & $0.07\pm0.09$ & $0.10\pm0.06$ & $0.10\pm0.05$ & $14.95\pm51.16$ & $15.13\pm51.11$ & $0.78\pm0.05$\\
& $M$ & $-0.16\pm0.50$ & $1.18\pm0.58$ & $1.63\pm0.72$ & $-0.01\pm0.12$ & $0.16\pm0.13$ & $0.23\pm0.15$\\
$B$ & $K$ & $0.00\pm0.16$ & $0.38\pm0.13$ & $0.62\pm0.22$ & $0.01\pm0.16$ & $0.16\pm0.15$ & $0.28\pm0.24$\\
& $e$ & $0.06\pm0.09$ & $0.10\pm0.06$ & $0.10\pm0.04$ & $15.01\pm49.58$ & $15.20\pm49.22$ & $0.77\pm0.05$\\
& $M$ & $-0.20\pm0.51$ & $1.14\pm0.50$ & $1.60\pm0.62$ & $0.01\pm0.11$ & $0.15\pm0.12$ & $0.23\pm0.14$\\
\hline\hline
\rule{0pt}{3ex} 
Strategy&Parameter&&Absolute&&&Relative&\\
&&\multicolumn{3}{c}{\rule{0.3cm}{0.0cm}\rule{5.8cm}{0.03cm}}&\multicolumn{3}{c}{\rule{0.3cm}{0.0cm}\rule{5.2cm}{0.03cm}}\\
&&Bias&Accuracy&Precision&Bias&Accuracy&Precision\\
\rule{0pt}{3ex} 
$A_1$ & $K$ & $-11.38\pm28.78$ & $16.36\pm26.32$ & $13.14\pm13.74$ & $-0.11\pm0.25$ & $0.15\pm0.22$ & $0.16\pm0.09$\\
& $e$ & $0.04\pm0.09$ & $0.06\pm0.07$ & $0.06\pm0.03$ & $2.93\pm5.18$ & $3.05\pm5.11$ & $0.81\pm0.39$\\
& $P$ & $-82.47\pm580.39$ & $335.58\pm483.62$ & $406.67\pm374.74$ & $-0.01\pm0.21$ & $0.16\pm0.15$ & $0.61\pm1.01$\\
& $M$ & $-242.48\pm763.63$ & $422.76\pm681.17$ & $339.97\pm438.22$ & $-0.12\pm0.28$ & $0.80\pm1.10$ & $0.20\pm0.12$\\
$A_2$ & $K$ & $-11.55\pm27.10$ & $16.05\pm24.72$ & $13.72\pm13.14$ & $-0.11\pm0.23$ & $0.15\pm0.21$ & $0.16\pm0.09$\\
& $e$ & $0.01\pm0.04$ & $0.03\pm0.03$ & $0.05\pm0.03$ & $2.61\pm5.70$ & $2.74\pm5.64$ & $0.88\pm0.42$\\
& $P$ & $-75.99\pm570.33$ & $334.29\pm468.41$ & $377.98\pm391.85$ & $-0.02\pm0.20$ & $0.14\pm0.15$ & $0.21\pm0.11$\\
& $M$ & $-255.13\pm728.48$ & $421.20\pm647.21$ & $331.58\pm395.52$ & $-0.11\pm0.27$ & $0.74\pm1.11$ & $0.19\pm0.12$\\
$B$ & $K$ & $-9.69\pm25.37$ & $13.79\pm23.41$ & $12.35\pm12.59$ & $-0.08\pm0.21$ & $0.13\pm0.19$ & $0.16\pm0.10$\\
& $e$ & $0.02\pm0.03$ & $0.03\pm0.02$ & $0.06\pm0.03$ & $2.29\pm4.77$ & $2.40\pm4.71$ & $0.82\pm0.32$\\
& $P$ & $-41.40\pm558.24$ & $301.71\pm474.52$ & $384.89\pm389.36$ & $0.02\pm0.21$ & $0.14\pm0.15$ & $0.45\pm0.71$\\
& $M$ & $-206.87\pm697.44$ & $369.75\pm626.67$ & $310.89\pm431.26$ & $-0.08\pm0.24$ & $1.08\pm1.45$ & $0.22\pm0.15$\\
\hline\hline
\end{tabular}
\label{table:t4}
\end{table*}

\begin{table*}
\caption{Absolute and relative bias, accuracy and precision with which $\eta_1$, $\eta_2$, $\eta_3$, $\eta_4$ and $s$, are recovered, averaged over all simulations, for the three strategies. The absolute quantities with respect to $\eta_1$ and s are in units of m/s, in the case of $\eta_2$ and $\eta_3$ are in units of days, while $\eta_4$ is dimensionless.}
\centering
\begin{tabular}{c c c c c c c c}
\hline\hline
\rule{0pt}{3ex} 
Strategy&Parameter&&Absolute&&&Relative&\\
&&\multicolumn{3}{c}{\rule{0.3cm}{0.0cm}\rule{5.0cm}{0.03cm}}&\multicolumn{3}{c}{\rule{0.3cm}{0.0cm}\rule{4.7cm}{0.03cm}}\\
&&Bias&Accuracy&Precision&Bias&Accuracy&Precision\\
\rule{0pt}{3ex} 
$A_1$ & $\eta_1$ & $0.20\pm1.32$ & $0.85\pm1.13$ & $0.79\pm0.74$ & $0.36\pm1.84$ & $0.80\pm1.72$ & $0.57\pm0.16$\\
& $\eta_2$ & $3.21\pm27.17$ & $22.21\pm16.04$ & $16.44\pm5.79$ & $0.62\pm1.31$ & $0.94\pm1.11$ & $0.43\pm0.04$\\
& $\eta_3$ & $-0.01\pm7.94$ & $6.23\pm4.93$ & $0.88\pm0.94$ & $0.16\pm0.66$ & $0.47\pm0.49$ & $0.05\pm0.05$\\
& $\eta_4$ & $0.47\pm0.16$ & $0.47\pm0.16$ & $0.59\pm0.06$ & $0.68\pm0.24$ & $0.69\pm0.24$ & $0.53\pm0.05$\\
& $s$ & $0.54\pm0.77$ & $0.60\pm0.73$ & $0.61\pm0.63$ & $2.30\pm3.20$ & $2.41\pm3.13$ & $0.65\pm0.15$\\
$A_2$ & $\eta_1$ & $-0.70\pm0.74$ & $0.65\pm0.50$ & $0.62\pm0.23$ & $-0.06\pm0.54$ & $0.48\pm0.32$ & $0.61\pm0.12$\\
& $\eta_2$ & $3.87\pm27.73$ & $22.50\pm16.70$ & $16.61\pm5.74$ & $0.65\pm1.35$ & $0.96\pm1.15$ & $0.43\pm0.04$\\
& $\eta_3$ & $0.00\pm7.96$ & $6.22\pm4.97$ & $0.90\pm0.96$ & $0.16\pm0.66$ & $0.47\pm0.49$ & $0.05\pm0.05$\\
& $\eta_4$ & $0.51\pm0.16$ & $0.51\pm0.16$ & $0.62\pm0.04$ & $0.74\pm0.25$ & $0.74\pm0.25$ & $0.53\pm0.06$\\
& $s$ & $0.55\pm0.42$ & $0.60\pm0.36$ & $0.52\pm0.17$ & $2.37\pm2.58$ & $2.44\pm2.52$ & $0.58\pm0.11$\\
$B$ & $\eta_1$ & $-0.40\pm0.55$ & $0.65\pm0.42$ & $0.60\pm0.23$ & $-0.20\pm0.38$ & $0.47\pm0.21$ & $0.71\pm0.11$\\
& $\eta_2$ & $2.66\pm26.31$ & $21.35\pm15.60$ & $16.94\pm5.80$ & $0.59\pm1.28$ & $0.90\pm1.09$ & $0.44\pm0.02$\\
& $\eta_3$ & $-0.01\pm7.93$ & $6.19\pm4.95$ & $0.97\pm1.05$ & $0.15\pm0.65$ & $0.46\pm0.48$ & $0.05\pm0.05$\\
& $\eta_4$ & $0.50\pm0.13$ & $0.50\pm0.13$ & $0.64\pm0.01$ & $0.73\pm0.21$ & $0.73\pm0.21$ & $0.54\pm0.04$\\
& $s$ & $0.83\pm0.50$ & $0.85\pm0.46$ & $0.48\pm0.16$ & $3.38\pm3.30$ & $3.41\pm3.27$ & $0.42\pm0.07$\\
\hline\hline
\end{tabular}
\label{table:t5}
\end{table*}

The estimation of $M$ is most dependent of $K$, but it is also contingent on the values for $e$, $P$ and the stellar mass. Thus, it is not straightforward to extrapolate results for $K$ to what would be expected with respect to $M$. In order to estimate $M$ one also needs to assume an inclination for the orbital plane. We will assume this to be known, and set it to $90^{\rm o}$, the same value assumed for all systems when the RV measurements were simulated. Although this situation is not realistic, it allows for a direct comparison between true and estimated planetary masses.

Overall, the two uniform-in-phase scheduling strategies, myopic, A2, and non-myopic, B, lead to very similar results. In the case of the transiting planets, the values estimated for $K$ and $M$ are significantly less biased, as well as more accurate and precise than those obtained through the random strategy, A1. However, there are no significant differences between the three strategies with respect to how well the true values of $e$ are recovered. Given that most of these are about $0.1$ or smaller, as can be seen in Figure \ref{fig:i1}, it is not surprising to find that all scheduling strategies lead to values around $0.1$ or smaller for the absolute bias, accuracy and precision, and thus much higher values for the relative counterparts to these quantities.

With respect to the detected non-transiting planets, all scheduling strategies lead to the acquisition of similar amounts of information about the true values of $K$, $e$, $P$ and $M$. This is not surprising, given that none of the strategies was designed with the aim of detecting such planets. The same happens with respect to the parameters associated with the Gaussian process model that is used to describe the stellar activity induced RV variations. Interestingly, the expected values for $K$, as well as for the mass, $M$, derived for the detected non-transiting planets given the simulated datasets are typically smaller by a factor of about 10\% with respect to the true values. However, the expected values for the orbital period, $P$, are essentially unbiased with respect to the true values. As expected, in particular given the discussion in Subsection 3.2, the most important factors affecting the detection probability of a non-transiting planet in our simulations are the number of RV measurements available and the value of $K/\sigma_{\rm act}$. In strategies A2 and B, the former is almost always 22, which is enough for the detection of 8 non-transiting planets in all simulations. But the non-transiting planet in system 40, which has a significantly lower value for $K/\sigma_{\rm act}$, just 5.6, is only detected 50\% and 70\% of the times in strategies A2 and B, respectively. This suggests that for $N_{\rm RV}$ around 22 only planets with $K$ in excess of roughly $5.6\times\sigma_{\rm act}$ can be detected with a probability greater than 50\%. On the other hand, although the planets in systems 11 and 45, which have very similar values for $K/\sigma_{\rm act}$ (close to 68), are always detected in the simulations of strategies A2 and B, they are only detected 70\% and 60\% of the times, respectively, under strategy A1. This is due to $N_{\rm RV}$ falling below 12 in the simulations were detection did not occur.

In the case of the transiting planets, all the distributions associated with the bias, accuracy and precision are positively skewed, except those for the bias and precision with respect to $e$ for which the skew is negative. The non-zero mean and positive skew in the distribution of the bias for $K$ and $M$, seems to be the result of the existence of undetected (non-transiting) planets. The mean bias gets closer to zero and the skew greatly diminishes, if only datasets whose analysis lead to the detection  of all non-transiting planets in the associated systems are considered in the calculation of these statistics (and the opposite occurs for the other systems). Interestingly, in the case of the uniform-in-phase strategies, the sampling of the phase-curves of the transiting planets seems to be so close to optimal in terms of information gathering, that even in the presence of undetected (non-transiting) planets the bias is very close to zero and the skew small.

The differences between the results obtained for each scheduling strategy, regarding both the transiting and non-transiting planets, should increase as the average number of possible RV measurements per star, $N_{\rm RV}$, decreases, and vice-versa. For example, if this number was about half of what was assumed, i.e. around 10, we would still expect strategy B, as well as A2 to a lesser extent, to yield fairly strong constraints on the masses and orbital parameters of the transiting planets, but it would be hard to detect any non-transiting planet. On the contrary, in this situation, strategy A1 would probably fail to deliver reliable constraints for the transiting planets around the least observed stars, but some non-transiting planets would end up being detected around the most observed stars. Although the mean $N_{\rm RV}$ is, by construction, exactly the same for all the scheduling strategies, the associated standard deviation is $8.39$ for strategy A1, while only $1.44$ for A2 and $1.05$ for B.

In the Appendix, we present the results obtained by assuming the RV variations induced by stellar activity are uncorrelated, and can be described as Gaussian white noise. As expected, given that none of the three scheduling strategies considered relies on the assumed model for such variations to decide on the best schedule, the conclusions that can be drawn are very similar to the ones just described.

\section{Conclusions}

We implemented three different scheduling strategies for the ESPRESSO GTO allocated to radial velocity follow-up of TOIs. Our main objective was to compare a novel uniform-in-phase scheduling algorithm with a random scheduler, and determine whether a non-myopic implementation of the former offered any advantage with respect to the more common myopic way. The scheduling strategies were compared with respect to the amount of information gathered about the masses and orbital parameters of all planets in the TOIs host systems. In particular, we considered a sample of 50 TESS target stars, with simulated planetary systems containing at least one transiting planet with a radius below $4R_{\oplus}$ \citep{Barclay2018}. 

We found that both uniform-in-phase scheduling strategies lead to an unbiased (at the level of 1\%) measurement of the masses of the transiting planets, while keeping the average accuracy and precision around 16\% and 23\%, respectively. This is significantly better than what can be achieved with random scheduling, which does not only lead to more biased (about 2\%) estimates of the mass of the simulated TOIs, but also to less accurate and precise estimates, respectively about 19\% and 28\% on average. The number of non-transiting planets detected is similar for all the scheduling strategies considered, as well as the bias, accuracy and precision with which their masses and orbital parameters are recovered.
 
Although we have not found any significant difference between the results obtained with the two uniform-in-phase scheduling strategies, myopic and non-myopic, this may be due to an assumed timespan for the observations (3 years) that is much larger than the orbital periods of the target transiting planets (below 50 days). As this difference decreases, a myopic scheduling strategy should lead to increasingly larger deviations with respect to uniform sampling of the phase curves, given that less than optimal choices early on become more difficult to compensate later in the observation schedule.

\section*{Acknowledgements}
We thank Nuno Santos for insightful discussions. We acknowledge the excellent open-source acebayes R package made available to the community by Antony Overstall. This work was supported by Funda\c{c}\~{a}o para a Ci\^{e}ncia e a Tecnologia (FCT) through national funds (PIDDAC) and the research grants UID/FIS/04434/2019, UIDB/04434/2020 and UIDP/04434/2020. This work was also supported by FCT through national funds (PTDC/FIS-AST/28953/2017, PTDC/FIS-AST/32113/2017) and by FEDER - Fundo Europeu de Desenvolvimento Regional through COMPETE2020 - Programa Operacional Competitividade e Internacionaliza\c{c}\~{a}o (POCI-01-0145-FEDER-028953, POCI-01-0145-FEDER-032113).

\section*{Data availability}
All the data underlying this article will be shared on request to the corresponding author.



\bibliographystyle{mnras}
\bibliography{biblio2} 

\begin{thebibliography}{}
\makeatletter
\relax
\def\mn@urlcharsother{\let\do\@makeother \do\$\do\&\do\#\do\^\do\_\do\%\do\~}
\def\mn@doi{\begingroup\mn@urlcharsother \@ifnextchar [ {\mn@doi@}
  {\mn@doi@[]}}
\def\mn@doi@[#1]#2{\def\@tempa{#1}\ifx\@tempa\@empty \href
  {http://dx.doi.org/#2} {doi:#2}\else \href {http://dx.doi.org/#2} {#1}\fi
  \endgroup}
\def\mn@eprint#1#2{\mn@eprint@#1:#2::\@nil}
\def\mn@eprint@arXiv#1{\href {http://arxiv.org/abs/#1} {{\tt arXiv:#1}}}
\def\mn@eprint@dblp#1{\href {http://dblp.uni-trier.de/rec/bibtex/#1.xml}
  {dblp:#1}}
\def\mn@eprint@#1:#2:#3:#4\@nil{\def\@tempa {#1}\def\@tempb {#2}\def\@tempc
  {#3}\ifx \@tempc \@empty \let \@tempc \@tempb \let \@tempb \@tempa \fi \ifx
  \@tempb \@empty \def\@tempb {arXiv}\fi \@ifundefined
  {mn@eprint@\@tempb}{\@tempb:\@tempc}{\expandafter \expandafter \csname
  mn@eprint@\@tempb\endcsname \expandafter{\@tempc}}}

\bibitem[\protect\citeauthoryear{Angus, Morton, Aigrain, Foreman-Mackey  \&
  Rajpaul}{Angus et~al.}{2018}]{Angus2018}
Angus R.,  Morton T.,  Aigrain S.,  Foreman-Mackey D.,   Rajpaul V.,  2018,
  Monthly Notices of the Royal Astronomical Society, 474, 2094

\bibitem[\protect\citeauthoryear{Barclay, Pepper  \& Quintana}{Barclay
  et~al.}{2018}]{Barclay2018}
Barclay T.,  Pepper J.,   Quintana E.~V.,  2018, The Astrophysical Journal
  Supplement Series, 239, 2

\bibitem[\protect\citeauthoryear{Batalha, Kempton  \& Mbarek}{Batalha
  et~al.}{2017}]{BKM2017}
Batalha N.~E.,  Kempton E. M.-R.,   Mbarek R.,  2017, The Astrophysical Journal
  Letters, 836, L5

\bibitem[\protect\citeauthoryear{Brewer \& Donovan}{Brewer \&
  Donovan}{2015}]{BD2015}
Brewer B.~J.,  Donovan C.~P.,  2015, Monthly Notices of the Royal Astronomical
  Society, 448, 3206

\bibitem[\protect\citeauthoryear{Brown et~al.,}{Brown et~al.}{2016}]{Gaia2016}
Brown A.~G.,  et~al., 2016, Astronomy \& Astrophysics, 595, A2

\bibitem[\protect\citeauthoryear{Brown et~al.,}{Brown et~al.}{2018}]{Gaia2018}
Brown A.,  et~al., 2018, Astronomy \& astrophysics, 616, A1

\bibitem[\protect\citeauthoryear{Burt, Holden, Wolfgang  \& Bouma}{Burt
  et~al.}{2018}]{Burt2018}
Burt J.,  Holden B.,  Wolfgang A.,   Bouma L.,  2018, The Astronomical Journal,
  156, 255

\bibitem[\protect\citeauthoryear{Cameron}{Cameron}{2018}]{Cameron2018}
Cameron A.~C.,  2018, Handbook of Exoplanets, pp 1791--1799

\bibitem[\protect\citeauthoryear{Cegla}{Cegla}{2019}]{Cegla2019}
Cegla H.,  2019, Geosciences, 9, 114

\bibitem[\protect\citeauthoryear{Cegla, Stassun, Watson, Bastien  \&
  Pepper}{Cegla et~al.}{2013}]{Cegla2014}
Cegla H.,  Stassun K.,  Watson C.,  Bastien F.,   Pepper J.,  2013, The
  Astrophysical Journal, 780, 104

\bibitem[\protect\citeauthoryear{Chen \& Kipping}{Chen \&
  Kipping}{2016}]{CK2017}
Chen J.,  Kipping D.,  2016, The Astrophysical Journal, 834, 17

\bibitem[\protect\citeauthoryear{Cloutier et~al.,}{Cloutier
  et~al.}{2017}]{Cloutier2017}
Cloutier R.,  et~al., 2017, Astronomy \& Astrophysics, 608, A35

\bibitem[\protect\citeauthoryear{Cloutier, Doyon, Bouchy  \&
  H{\'e}brard}{Cloutier et~al.}{2018}]{Cloutier2018}
Cloutier R.,  Doyon R.,  Bouchy F.,   H{\'e}brard G.,  2018, The Astronomical
  Journal, 156, 82

\bibitem[\protect\citeauthoryear{Dorn, Khan, Heng, Connolly, Alibert, Benz  \&
  Tackley}{Dorn et~al.}{2015}]{Dorn2015}
Dorn C.,  Khan A.,  Heng K.,  Connolly J.~A.,  Alibert Y.,  Benz W.,   Tackley
  P.,  2015, Astronomy \& Astrophysics, 577, A83

\bibitem[\protect\citeauthoryear{Dorn, Bower  \& Rozel}{Dorn
  et~al.}{2017}]{DBR2018}
Dorn C.,  Bower D.~J.,   Rozel A.,  2017, Handbook of Exoplanets, pp 1--25

\bibitem[\protect\citeauthoryear{Dressing \& Charbonneau}{Dressing \&
  Charbonneau}{2015}]{DC2015}
Dressing C.~D.,  Charbonneau D.,  2015, The Astrophysical Journal, 807, 45

\bibitem[\protect\citeauthoryear{Dumusque}{Dumusque}{2016}]{Dumusque2016}
Dumusque X.,  2016, Astronomy \& Astrophysics, 593, A5

\bibitem[\protect\citeauthoryear{Dumusque, Udry, Lovis, Santos  \&
  Monteiro}{Dumusque et~al.}{2011}]{Dumusque2011}
Dumusque X.,  Udry S.,  Lovis C.,  Santos N.~C.,   Monteiro M.,  2011,
  Astronomy \& Astrophysics, 525, A140

\bibitem[\protect\citeauthoryear{Faria, Haywood, Brewer, Figueira, Oshagh,
  Santerne  \& Santos}{Faria et~al.}{2016}]{Faria2016}
Faria J.,  Haywood R.,  Brewer B.,  Figueira P.,  Oshagh M.,  Santerne A.,
  Santos N.,  2016, Astronomy \& Astrophysics, 588, A31

\bibitem[\protect\citeauthoryear{Faria, Santos, Figueira  \& Brewer}{Faria
  et~al.}{2018}]{FSFB2018}
Faria J.~P.,  Santos N.~C.,  Figueira P.,   Brewer B.~J.,  2018, Journal of
  Open Source Software, 3, 487

\bibitem[\protect\citeauthoryear{Faria et~al.,}{Faria et~al.}{2020}]{Faria2020}
Faria J.,  et~al., 2020, Astronomy \& Astrophysics, 635, A13

\bibitem[\protect\citeauthoryear{Feroz \& Hobson}{Feroz \&
  Hobson}{2013}]{FH2013}
Feroz F.,  Hobson M.,  2013, Monthly Notices of the Royal Astronomical Society,
  437, 3540

\bibitem[\protect\citeauthoryear{Feroz, Balan  \& Hobson}{Feroz
  et~al.}{2011}]{FBH2011}
Feroz F.,  Balan S.,   Hobson M.,  2011, Monthly Notices of the Royal
  Astronomical Society, 415, 3462

\bibitem[\protect\citeauthoryear{Ford}{Ford}{2008}]{Ford2008}
Ford E.~B.,  2008, The Astronomical Journal, 135, 1008–

\bibitem[\protect\citeauthoryear{Fressin, Torres, Charbonneau, Bryson,
  Christiansen  \& Dressing}{Fressin et~al.}{2013}]{Fressin2013}
Fressin F.,  Torres G.,  Charbonneau D.,  Bryson S.~T.,  Christiansen J.,
  Dressing C.~D.,  2013, The Astrophysical Journal, 766, 81

\bibitem[\protect\citeauthoryear{Gladman}{Gladman}{1993}]{Gladman1993}
Gladman B.,  1993, Icarus, 106, 247

\bibitem[\protect\citeauthoryear{Haywood et~al.,}{Haywood
  et~al.}{2014}]{Haywood2014}
Haywood R.,  et~al., 2014, Monthly Notices of the Royal Astronomical Society,
  443, 2517

\bibitem[\protect\citeauthoryear{Hees, Dehghanfar, Do, Ghez, Martinez, Campbell
   \& Lu}{Hees et~al.}{2019}]{Hees2019}
Hees A.,  Dehghanfar A.,  Do T.,  Ghez A.,  Martinez G.,  Campbell R.,   Lu J.,
   2019, The Astrophysical Journal, 880, 87

\bibitem[\protect\citeauthoryear{Herman, Zhu  \& Wu}{Herman
  et~al.}{2019}]{HZW2019}
Herman M.~K.,  Zhu W.,   Wu Y.,  2019, The Astronomical Journal, 157, 248

\bibitem[\protect\citeauthoryear{Jeffreys}{Jeffreys}{1998}]{Jeffreys1998}
Jeffreys H.,  1998, The Theory of Probability 3rd Edition (Reissue).
Oxford: Oxford University Press) ISBN--13: 9780198503682

\bibitem[\protect\citeauthoryear{Kanodia, Wolfgang, Stefansson, Ning  \&
  Mahadevan}{Kanodia et~al.}{2019}]{Kanodia2019}
Kanodia S.,  Wolfgang A.,  Stefansson G.~K.,  Ning B.,   Mahadevan S.,  2019,
  The Astrophysical Journal, 882, 38

\bibitem[\protect\citeauthoryear{Kass \& Raftery}{Kass \&
  Raftery}{1995}]{KR1995}
Kass R.~E.,  Raftery A.~E.,  1995, Journal of the American Statistical
  Association, 90, 773

\bibitem[\protect\citeauthoryear{Kipping}{Kipping}{2013}]{Kipping2013}
Kipping D.~M.,  2013, Monthly Notices of the Royal Astronomical Society:
  Letters, 434, L51

\bibitem[\protect\citeauthoryear{Kipping}{Kipping}{2014}]{Kipping2014}
Kipping D.~M.,  2014, Monthly Notices of the Royal Astronomical Society, 444,
  2263

\bibitem[\protect\citeauthoryear{Korhonen, Andersen, Piskunov, Hackman,
  Juncher, J{\"a}rvinen  \& J{\o}rgensen}{Korhonen et~al.}{2015}]{Korhonen2015}
Korhonen H.,  Andersen J.,  Piskunov N.,  Hackman T.,  Juncher D.,
  J{\"a}rvinen S.,   J{\o}rgensen U.~G.,  2015, Monthly Notices of the Royal
  Astronomical Society, 448, 3038

\bibitem[\protect\citeauthoryear{Kumaraswamy}{Kumaraswamy}{1980}]{Kumaraswamy1980}
Kumaraswamy P.,  1980, Journal of Hydrology, 46, 79

\bibitem[\protect\citeauthoryear{Kushniruk, Schirmer  \& Bensby}{Kushniruk
  et~al.}{2017}]{KSB2017}
Kushniruk I.,  Schirmer T.,   Bensby T.,  2017, Astronomy \& Astrophysics, 608,
  A73

\bibitem[\protect\citeauthoryear{Landoni, Romano, Vercellone, Kn{\"o}dlseder,
  Bianco, Tavecchio  \& Corina}{Landoni et~al.}{2019}]{Landoni2019}
Landoni M.,  Romano P.,  Vercellone S.,  Kn{\"o}dlseder J.,  Bianco A.,
  Tavecchio F.,   Corina A.,  2019, The Astrophysical Journal Supplement
  Series, 240, 32

\bibitem[\protect\citeauthoryear{Lopez-Morales et~al.,}{Lopez-Morales
  et~al.}{2016}]{Morales2016}
Lopez-Morales M.,  et~al., 2016, The Astronomical Journal, 152, 204

\bibitem[\protect\citeauthoryear{Loredo, Berger, Chernoff, Clyde  \&
  Liu}{Loredo et~al.}{2012}]{Loredo2012}
Loredo T.~J.,  Berger J.~O.,  Chernoff D.~F.,  Clyde M.~A.,   Liu B.,  2012,
  Statistical Methodology, 9, 101

\bibitem[\protect\citeauthoryear{McQuillan, Mazeh  \& Aigrain}{McQuillan
  et~al.}{2014}]{McQuillan2014}
McQuillan A.,  Mazeh T.,   Aigrain S.,  2014, The Astrophysical Journal
  Supplement Series, 211, 24

\bibitem[\protect\citeauthoryear{Montet}{Montet}{2018}]{Montet2018}
Montet B.~T.,  2018, Research Notes of the American Astronomical Society, 2, 28

\bibitem[\protect\citeauthoryear{Ning, Wolfgang  \& Ghosh}{Ning
  et~al.}{2018}]{NWG2018}
Ning B.,  Wolfgang A.,   Ghosh S.,  2018, The Astrophysical Journal, 869, 5

\bibitem[\protect\citeauthoryear{Overstall \& Woods}{Overstall \&
  Woods}{2017}]{OW2017}
Overstall A.~M.,  Woods D.~C.,  2017, Technometrics, 59

\bibitem[\protect\citeauthoryear{Overstall, Woods  \& Adamou}{Overstall
  et~al.}{2017}]{OWA2017}
Overstall A.,  Woods D.,   Adamou M.,  2017, arXiv:1705.08096

\bibitem[\protect\citeauthoryear{Pepe et~al.,}{Pepe et~al.}{2013}]{Pepe2013}
Pepe F.,  et~al., 2013, The Messenger, 153, 6

\bibitem[\protect\citeauthoryear{Pepe et~al.,}{Pepe et~al.}{2014}]{Pepe2014}
Pepe F.,  et~al., 2014, Astronomische Nachrichten, 335, 8

\bibitem[\protect\citeauthoryear{Pepe et~al.,}{Pepe et~al.}{2020}]{Pepe2020}
Pepe F.,  et~al., 2020, arXiv:2010.00316

\bibitem[\protect\citeauthoryear{Perryman}{Perryman}{2018}]{Perryman2018}
Perryman M.,  2018, The Exoplanet Handbook.
Cambridge University Press

\bibitem[\protect\citeauthoryear{Pronzato}{Pronzato}{2017}]{Pronzato2017}
Pronzato L.,  2017, Journal de la Soci\'{e}t\'{e} Française de Statistique, pp
  7--36

\bibitem[\protect\citeauthoryear{Rajpaul}{Rajpaul}{2017}]{Rajpaul2017}
Rajpaul V.~M.,  2017, PhD thesis.
University of Oxford

\bibitem[\protect\citeauthoryear{Ricker et~al.,}{Ricker et~al.}{2016}]{R2016}
Ricker G.~R.,  et~al., 2016, in Space Telescopes and Instrumentation 2016:
  Optical, Infrared, and Millimeter Wave. p. 99042B

\bibitem[\protect\citeauthoryear{Suissa, Chen  \& Kipping}{Suissa
  et~al.}{2018}]{SCK2018}
Suissa G.,  Chen J.,   Kipping D.,  2018, Monthly Notices of the Royal
  Astronomical Society, 476, 2613–2620

\bibitem[\protect\citeauthoryear{Tayar, Stassun  \& Corsaro}{Tayar
  et~al.}{2019}]{TSC2018}
Tayar J.,  Stassun K.~G.,   Corsaro E.,  2019, The Astrophysical Journal, 883,
  195

\bibitem[\protect\citeauthoryear{Van~Eylen et~al.,}{Van~Eylen
  et~al.}{2019}]{V2019}
Van~Eylen V.,  et~al., 2019, The Astronomical Journal, 157, 61

\bibitem[\protect\citeauthoryear{Winn \& Fabrycky}{Winn \&
  Fabrycky}{2015}]{WF2015}
Winn J.~N.,  Fabrycky D.~C.,  2015, Annual Review of Astronomy and
  Astrophysics, 53, 409

\bibitem[\protect\citeauthoryear{Wolfgang, Rogers  \& Ford}{Wolfgang
  et~al.}{2016}]{WRF2016}
Wolfgang A.,  Rogers L.~A.,   Ford E.~B.,  2016, The Astrophysical Journal,
  825, 19

\makeatother
\end{thebibliography}



\appendix

\section{Appendix A}

\subsection{Results obtained assuming only Gaussian white noise}

\begin{figure*}
\includegraphics[width=\linewidth]{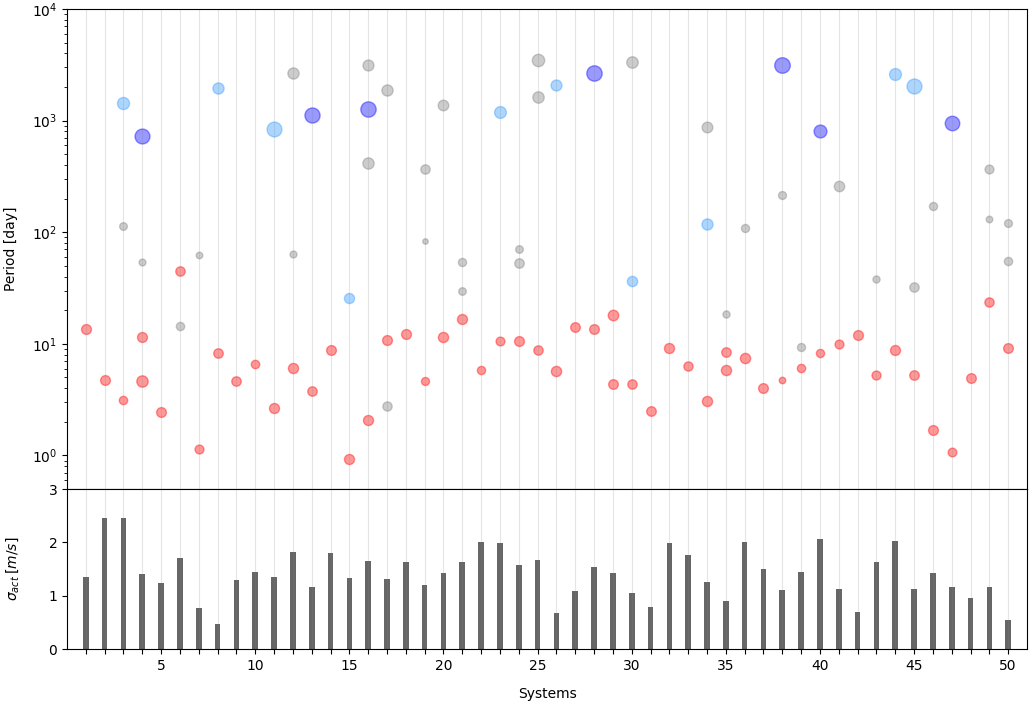}
\caption{The upper panel shows the planetary system architecture as a function of orbital period. Light red circles represent transiting planets, while non-transiting planets are represented by grey circles, when never detected, light blue circles, when detected at least once, and violet circles, when always detected, with respect to the $3\times10$ simulations carried out for the three scheduling strategies under the assumption of non-correlated stellar activity induced RV variations. The planets coloured light blue were detected with strategies [A1, A2, B], respectively, the following number of times: system 3 [5, 2, 0]; system 8 [2, 1, 7]; system 11 [8, 10, 10]; system 15 [2, 1, 1]; system 23 [1, 1, 3]; system 26 [0, 2, 1]; system 30 [0, 1, 0]; system 34 [1, 0, 1]; system 44 [0, 1, 2]; system 45  [9, 10, 10]. The size of each circle is proportional to the mass of the respective planet. In the lower panel, the amplitude of the stellar activity induced RV variations, $\sigma_{\rm act}$, is represented for each star in our sample. All stars and their associated planetary systems are identified by an incremental number where 1 corresponds to the lowest TESS ID number and 50 to the highest TESS ID number in our sample.}

\label{fig:d_wn}
\end{figure*}

Here we present the results of the analysis of the RV datasets generated assuming the stellar activity induced RV variations are akin to Gaussian white noise, and compare them with the results previously discussed.

In Figure \ref{fig:d_wn} we highlight the non-transiting planets that are never detected, sometimes detected or always detected, across all simulations and for all three scheduling strategies. Averaging over the 10 simulations per strategy, a total of $9.8\pm0.6$, $9.9\pm0.8$ and $10.5\pm1.2$ non-transiting planets are detected using strategies A1, A2 and B, respectively, out of the 50 that we simulated orbiting our sample of stars. As before, these numbers are very similar, and the differences not significant given the variation seen across the simulations. They are also 15 to 20\% higher than those obtained for the datasets with correlated stellar activity noise. This was expected, given that it is more difficult to disentangle correlated noise than uncorrelated noise from a signal.

In Table \ref{table:t_wn}, the absolute and relative bias, accuracy and precision with which $K$, $e$ and $M$ are recovered, averaged over all simulations and either all transiting planets (upper panel) or all detected non-transiting planets (lower panel), are shown for the three strategies. In the lower panel the same quantities are shown with respect to the orbital period, $P$, of the detected non-transiting planets.

\begin{table*}
\caption{In the upper panel it is shown the absolute and relative bias, accuracy and precision with which $K$, $e$ and mass, $M$, are recovered, averaged over all transiting planets and simulations, for the three strategies. The uncertainties provided are standard deviations, and characterise the dispersion of such values taking into account all transiting planets. The same quantities, as well as the orbital period, $P$, are provided in the lower panel with respect to all detected non-transiting planets. The absolute quantities with respect to $K$, $M$ and $P$ are in units of m/s, $M_{\odot}$ and days, respectively.}
\centering
\begin{tabular}{c c c c c c c c}
\hline\hline
\rule{0pt}{3ex} 
Strategy&Parameter&&Absolute&&&Relative&\\
&&\multicolumn{3}{c}{\rule{0.3cm}{0.0cm}\rule{5.8cm}{0.03cm}}&\multicolumn{3}{c}{\rule{0.3cm}{0.0cm}\rule{5.2cm}{0.03cm}}\\
&&Bias&Accuracy&Precision&Bias&Accuracy&Precision\\
\rule{0pt}{3ex} 
$A_1$ & $K$ & $0.25\pm0.36$ & $0.52\pm0.32$ & $0.90\pm0.54$ & $0.12\pm0.21$ & $0.21\pm0.20$ & $0.31\pm0.21$\\
& $e$ & $0.10\pm0.08$ & $0.11\pm0.06$ & $0.11\pm0.05$ & $14.39\pm44.04$ & $14.46\pm44.01$ & $0.70\pm0.10$\\
& $M$ & $0.40\pm0.90$ & $1.43\pm0.85$ & $2.27\pm1.59$ & $0.07\pm0.17$ & $0.19\pm0.17$ & $0.27\pm0.15$\\
$A_2$ & $K$ & $0.11\pm0.21$ & $0.44\pm0.18$ & $0.63\pm0.21$ & $0.06\pm0.15$ & $0.18\pm0.14$ & $0.26\pm0.21$\\
& $e$ & $0.08\pm0.08$ & $0.10\pm0.06$ & $0.10\pm0.04$ & $14.71\pm46.15$ & $14.80\pm46.512$ & $0.70\pm0.10$\\
& $M$ & $0.08\pm0.63$ & $1.26\pm0.64$ & $1.67\pm0.67$ & $0.03\pm0.12$ & $0.17\pm0.12$ & $0.23\pm0.14$\\
$B$ & $K$ & $0.05\pm0.19$ & $0.39\pm0.13$ & $0.58\pm0.19$ & $0.05\pm0.14$ & $0.16\pm0.13$ & $0.25\pm0.20$\\
& $e$ & $0.08\pm0.07$ & $0.09\pm0.06$ & $0.10\pm0.04$ & $13.64\pm39.94$ & $13.48\pm39.92$ & $0.69\pm0.11$\\
& $M$ & $-0.05\pm0.55$ & $1.15\pm0.50$ & $1.56\pm0.61$ & $0.01\pm0.11$ & $0.15\pm0.11$ & $0.22\pm0.14$\\
\hline\hline
\rule{0pt}{3ex} 
Strategy&Parameter&&Absolute&&&Relative&\\
&&\multicolumn{3}{c}{\rule{0.3cm}{0.0cm}\rule{5.8cm}{0.03cm}}&\multicolumn{3}{c}{\rule{0.3cm}{0.0cm}\rule{5.2cm}{0.03cm}}\\
&&Bias&Accuracy&Precision&Bias&Accuracy&Precision\\
\rule{0pt}{3ex} 
$A_1$ & $K$ & $-8.56\pm25.04$ & $11.97\pm23.64$ & $10.04\pm11.64$ & $-0.01\pm0.28$ & $0.19\pm0.22$ & $0.24\pm0.15$\\
& $e$ & $0.04\pm0.06$ & $0.05\pm0.05$ & $0.09\pm0.05$ & $3.32\pm5.62$ & $3.41\pm5.57$ & $0.87\pm0.30$\\
& $P$ & $56.87\pm568.78$ & $352.88\pm451.90$ & $626.04\pm467.78$ & $0.15\pm0.37$ & $0.26\pm0.31$ & $0.78\pm1.07$\\
& $M$ & $-180.26\pm654.75$ & $306.02\pm606.78$ & $257.13\pm356.90$ & $0.01\pm0.35$ & $10.04\pm20.02$ & $0.32\pm0.19$\\
$A_2$ & $K$ & $-5.96\pm21.12$ & $9.32\pm19.87$ & $8.08\pm9.88$ & $0.24\pm0.67$ & $0.38\pm0.60$ & $0.28\pm0.23$\\
& $e$ & $0.04\pm0.07$ & $0.06\pm0.06$ & $0.08\pm0.05$ & $2.79\pm4.44$ & $2.91\pm4.37$ & $0.87\pm0.34$\\
& $P$ & $257.82\pm729.02$ & $481.58\pm605.37$ & $701.87\pm722.12$ & $0.20\pm0.37$ & $0.28\pm0.31$ & $0.88\pm2.24$\\
& $M$ & $-128.34\pm556.24$ & $244.17\pm516.14$ & $214.77\pm324.32$ & $0.32\pm0.84$ & $22.29\pm39.88$ & $0.36\pm0.26$\\
$B$ & $K$ & $-6.27\pm21.98$ & $10.03\pm20.55$ & $8.50\pm10.39$ & $0.19\pm0.55$ & $0.35\pm0.46$ & $0.26\pm0.20$\\
& $e$ & $0.05\pm0.06$ & $0.06\pm0.05$ & $0.08\pm0.05$ & $3.39\pm5.37$ & $3.48\pm5.31$ & $0.87\pm0.36$\\
& $P$ & $268.97\pm853.49$ & $518.99\pm731.88$ & $759.89\pm702.79$ & $0.22\pm0.44$ & $0.31\pm0.37$ & $1.04\pm1.99$\\
& $M$ & $-137.18\pm584.83$ & $263.35\pm540.07$ & $220.32\pm346.71$ & $0.27\pm0.73$ & $23.90\pm48.20$ & $0.34\pm0.22$\\
\hline\hline
\end{tabular}
\label{table:t_wn}
\end{table*}

Overall, the results are very similar to the ones previously obtained under the assumption of correlated stellar activity induced RV variations. However, now the non-myopic uniform-in-phase strategy, B, seems consistently, though only slightly, better on average than the myopic strategy, A2, in terms of the information recovered about the true values of $K$ and $M$ for the transiting planets. Again, both these strategies lead on average to significantly less biased, as well as more accurate and precise values for $K$ and $M$ than strategy A1, with little difference between the three strategies with respect to how well the true values of $e$ are recovered. 

Somewhat counter-intuitively, the estimates for the mass and orbital parameters of the non-transiting planets seem now to be on average significantly more biased, less accurate, and less precise, than the estimates for the same quantities previously obtained under the assumption of correlated stellar activity induced RV variations. This is due to the fact that most of the 15 to 20\% extra non-transiting planets that are now being detected have substantially lower values for $K$. And given that the average number of RV measurements per system is independent of its characteristics, significantly less information is recovered about their mass and orbital parameters. In turn, this brings down the information recovered about such quantities when averaged over all detected non-transiting planets, making the averaged bias, accuracy and precision obtained under the assumption of non-correlated stellar activity noise seem worse than when such noise is assumed correlated.



\bsp	
\label{lastpage}
\end{document}